\documentclass{article}

\usepackage{arxiv}

\usepackage[utf8]{inputenc} 
\usepackage[T1]{fontenc}    
\usepackage{hyperref}       
\usepackage{url}            
\usepackage{booktabs}       
\usepackage{amsfonts}       
\usepackage{nicefrac}       
\usepackage{microtype}      

\usepackage{amssymb}
\usepackage{amsmath}
\usepackage{verbatim}
\usepackage{multirow}
\usepackage{caption}
\usepackage{subcaption}
\usepackage{float}
\usepackage{graphicx}

\title{High Temporal Resolution Rainfall Runoff Modelling Using Long-Short-Term-Memory (LSTM) Networks}

\author{
  Wei Li \\
  Oden Institute\\
  The University of Texas at Austin\\
  Austin, TX 78712 \\
  \texttt{wei@ices.utexas.edu} \\
   \And
  Amin Kiaghadi \\
  Oden Institute\\
  The University of Texas at Austin\\
  Austin, TX 78712 \\
  \texttt{kiaghadi@utexa.edu} \\
   \AND
   Clint N. Dawson \\
   Oden Institute \\
   The University of Texas at Austin\\
   Austin, TX 78712 \\
   \texttt{clint@oden.utexas.edu} \\
}

\begin{document}
\maketitle

\begin{abstract}
Accurate and efficient models for rainfall runoff (RR) simulations are crucial for flood risk management. Most rainfall models in use today are process-driven; i.e. they solve either simplified empirical formulas or some variation of the St. Venant (shallow water) equations.  With the development of machine-learning techniques, we may now be able to emulate rainfall models using, for example, neural networks.
In this study, a data-driven RR model using a sequence-to-sequence Long-short-Term-Memory (LSTM) network was constructed.  The model was tested for a watershed in Houston, TX, known for severe flood events. The LSTM network's capability in learning long-term dependencies between the input and output of the network allowed modeling RR with high resolution in time (15 minutes). Using 10-years precipitation from 153 rainfall gages and river channel discharge data (more than 5.3 million data points), and by designing several numerical tests the developed model performance in predicting river discharge was tested. The model results were also compared with the output of a process-driven model Gridded Surface Subsurface Hydrologic Analysis (GSSHA). Moreover, physical consistency of the LSTM model was explored. The model results showed that the LSTM model was able to efficiently predict discharge and achieve good model performance. When compared  to GSSHA,  the  data-driven model was more efficient and robust in terms of prediction and calibration. 
Interestingly, the performance of the LSTM model improved (test Nash-Sutcliffe model efficiency from 0.666 to 0.942) when a selected subset of rainfall gages based on the model performance, were used as input instead of all rainfall gages.
\end{abstract}

\keywords{Hydrologic Analysis \and GSSHA \and Recurrent Neural Network \and Machine Learning}

\section{Introduction}
\label{intro}
Flooding is considered the leading cause of natural-disaster losses in the United States with an average annual damage of \$7.95 billion (1984 - 2013, adjusted to 2014 inflation) \cite{NCEI2018}.
Implementing flood management strategies without a reliable predictive rainfall runoff (RR) modeling framework is not possible. RR modeling, which aims at predicting the streamflow hydrograph from precipitation input, is intensively studied and used to support flood assessment \cite{zhijia2008rainfall,talchabhadel2015rainfall,nayak2005fuzzy,pappenberger2005cascading,duan1992effective,zarzar2018hydraulic}.
In addition, RR models are required to provide reliable discharge input for storm surge models when real-time data is not availed due to the absence of measuring gages. Such coupling is very important to simulate the cascading effects of storm surge, local runoff, and compound flooding in coastal areas.  

RR models can be categorized as process-driven and data-driven models \cite{yuan2018monthly,Lu2013,Kan2015}. While process-driven methods are composed of analytical and empirical formulae based on physical phenomena, data-driven models rely on interpolating and extrapolating data. During the past two decades, multiple process-driven hydrologic models such as Interconnected Channel and Pond Routing Model (ICPR), Hydrologic Engineering Center's River Analysis System (HEC-RAS), and Gridded Surface Subsurface Hydrologic Analysis (GSSHA) for RR simulation have been developed \cite{channelversion,Bedient2000,brunner1997hec,Downer2004}.

Although much progress has been made recently, even state of the art process-driven models like GSSHA rely on accurate meteorological input data that are changing due to human/natural activities, which adds difficulty to constructing a production-level model incipiently. Furthermore, accurate prediction of RR requires extensive calibration of the multiphysics model that is computationally expensive and requires intensive data availability and entry. In addition, using process-driven models make it more difficult to build a coupled coast flood prediction model that considers the coupled interactions of hurricane storm surge and associated RR \cite{Torres2015}. During a flood event, channel discharge information computed by RR model will be passed to a surge model as a flux boundary condition. Concurrently water surface elevation computed by the surge model will be enforced on the RR watershed boundary as a Dirichlet boundary condition. The overhead caused by this message passing between RR and storm surge model will affect the overall computational efficiency.

Recently using deep neural networks for real time flood prediction has been made possible by the increasing amount of collected hydrologic data. Antithetical to process-driven models, data-driven models such as artificial neural network (ANN) that have been widely applied to streamflow prediction, e.g. \cite{hsu1995artificial,DAWSON1998,Sudheer2002,hettiarachchi2005extrapolation,Srinivasulu2006,Kan2015,young2015prediction}, are more robust to meteorological data changes. This robustness is due to the nature of their training data, batched learning, and relatively inexpensive calibration process.
Due to its capability of modeling highly nonlinear relationships between input and output, the ANN model has generated promising results for RR simulation.

When it comes to time series data, standard feed forward neural network has its limitations. Feed forward neural networks are designed based on the assumption that the training and test examples (data points) are independent. Thus the entire state of the network is erased after processing each example \cite{van2018artificial}. 
This assumption is not desired when data points are inherently related. Moreover, to deal with time series data, a standard ANN model (feed-forward) would require choosing a fixed-sized sliding window over the dataset. Tuning the size of this sliding windows for the best predictive accuracy adds extra work to the model selection \cite{hsu1995artificial}. This limitation becomes more significant in flood assessment with finer time resolution (i.e. 15 minutes). In this case, long-term dependencies prevail due to the small time step size and cannot be learned by ANN because they are not captured within the fixed-sized time windows. 

More recently, a class of ANNs known as recurrent neural network (RNN), a deep learning algorithm, has attracted much attention and shown success in solving sequential problems such as machine translation, speech recognition, and handwriting recognition \cite{farabet2013learning}. Even though the idea of RNN was proposed in the 1980s \cite{nerrand1993neural}, the applications of RNN in hydrologic engineering are relatively more recent \cite{taver2015feed,johannet2015neural}. RNNs are networks with loops in them, allowing information to persist. RNN can exploit the sequential pattern in the data while preserving feed-forward NN's ability to model nonlinear relationship between input and output via cycles formed by the hidden nodes in the network \cite{Lipton2015}.
A standard RNN has very simple looping units, such as a single layer with hyperbolic tangent (tanh) activation.
To cope with the vanishing gradient challenge \cite{hochreiter1998vanishing} for standard RNN and learn longer-term dependencies in sequential data, long short-term memory (LSTM) based RNN systems have been developed \cite{gers1999learning}.
LSTM's success has encouraged groups to explore its capability in time series forecasting of river discharge \cite{Kratzert2018,yuan2018monthly,Mhammedi2016,Wu2018}.

All of the aforementioned LSTM models have been using both rainfall and flow at previous time steps to predict future flow.
Even though the prediction uncertainty associated with time series forecast using LSTM is not analytically available yet, studies \cite{yuan2018monthly,Mhammedi2016,Wu2018} have shown increasing error in predicting flow by the passage of time.
Despite some examples in using RNN for hydrologic modeling \cite{faur2008modelling}, the literature obviously lacks an LSTM model that predicts future river flow purely based on precipitation input to address this accumulative uncertainty problem. In addition, such a model, capable of simulating longer events only using precipitation data as input, is more desirable for flood management applications and dynamic coupling with surge models.

Ubiquitous as deep learning systems are, they are often criticized for their lack of interpretability. There are multiple motivations to interpret deep learning models \cite{doshi2017towards}, among which two aspects raise the most concerns: (1) How can the prediction be trusted when the model is not interpretable? (2) How to select input features (rainfall gages) for the model? These two questions are especially difficult to answer for LSTM models. Unlike some machine learning models that have clearly defined importance metrics such as the random forest algorithm \cite{breiman2001random}, there is no simple way to define such a metric and offer insights to the learned LSTM model. Besides, the complicated structure of LSTM unit adds more difficulty to understand the prediction. While the importance of the first question is obvious, the second question is equally important as it is found that scrutinized gage selection can improve the model predictions \cite{lindstrom1997development,anctil2006improvement}. Thus, it is critical to understand the data-driven models and justify their results based on the physical intuition in addition to making good predictions.

In this paper, an LSTM network was applied to build a data-driven model for streamflow prediction in a urban watershed on a 15-minute scale and compared it with a benchmark process-driven model (GSSHA). 
The objectives of this paper are i) to build a data-driven model for streamflow prediction using precipitation as the only input in an urban watershed by applying LSTM network, ii) to compare the prediction accuracy and efficiency of the developed model with a benchmark process-driven model (GSSHA) and observational data, iii) to evaluate to what extent the model results can be justified based on the physical characteristics of the modeled watershed, and iv) to propose a fast methodology to reduce the dimension of input data to the model through an efficient feature selection approach.

\section{Study Area and Data Acquisition}
\label{sec:1}
\begin{figure}
    \centering
    \includegraphics[width=0.8\textwidth]{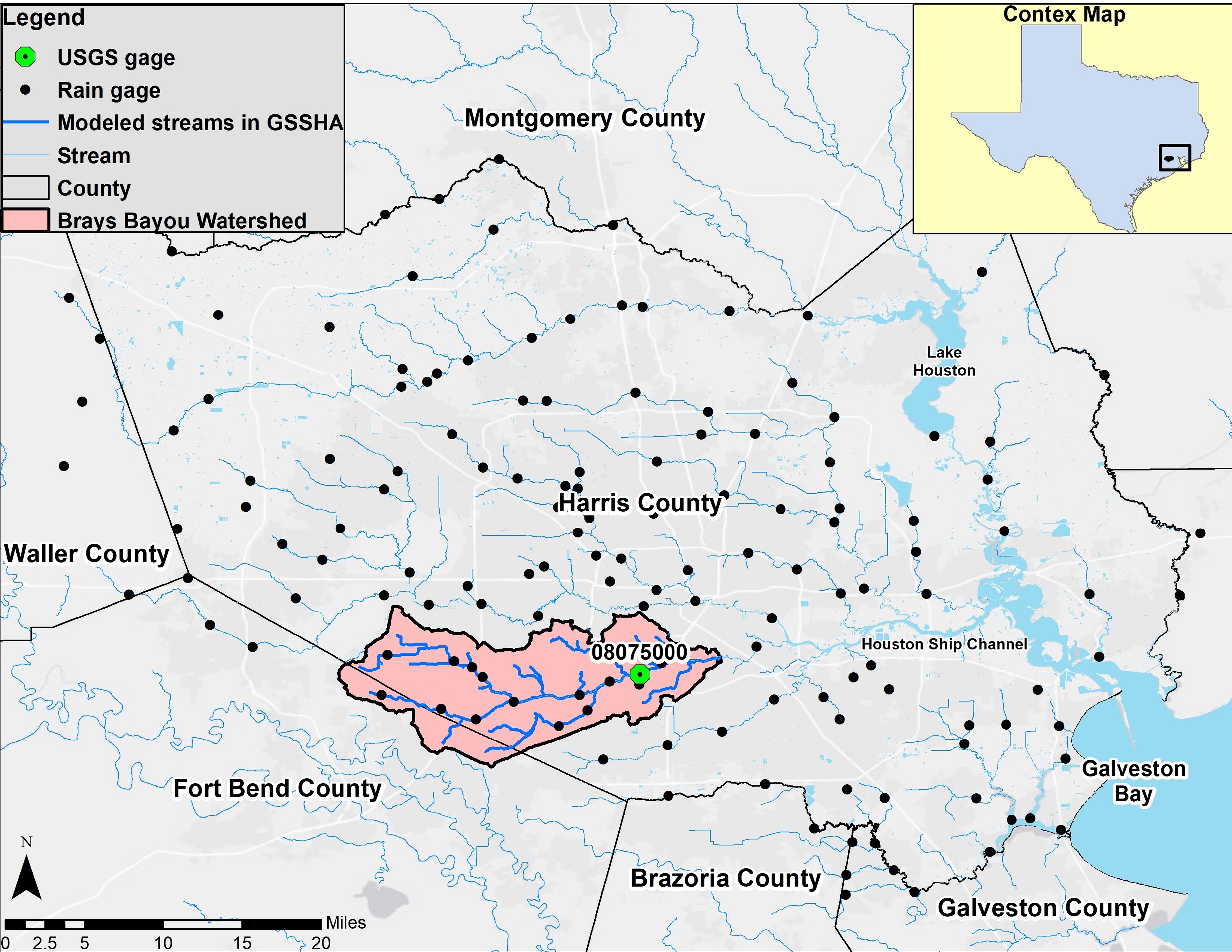}
    \caption{Study Area}
    \label{fig:study_area}
\end{figure}

Figure \ref{fig:study_area} shows the location of the Brays Bayou watershed, Brays Bayou and its tributaries located in southwest of Harris County and northeast of Fort Bend County, Texas was selected for this study. Brays Bayou drains freshwater from 329 square kilometers of a heavily urbanized and populated watershed and discharges into the Houston Ship Channel \cite{brays2019}. Brays Bayou has had a history of floods; just in the last 18 years Tropical Storm Allison (2001), Hurricane Ike (2008), the Memorial Day Flood (2015), the Tax Day Flood (2016), and Hurricane Harvey (2017) caused significant flooding and billions of dollars of property damage \cite{NCEI2018}.

15-minute precipitation data from 2007 to 2017 were compiled from 153 rainfall gages maintained by the Harris County Flood Control District (HCFCD) and 15 minutes flow data were obtained from the United States Geological Survey (USGS) gages \cite{harris2019}. Within the Brays Bayou watershed there are 16 rainfall gages and five flow gages. In this study only one freshwater gage located very close to the watershed outlet (see gage 08075000 in Figure \ref{fig:study_area}) was used to compile flow data for the purpose of training, calibration and validation.

Land elevation was extracted from the 10m resolution U.S. National Elevation Dataset (NED) in WMS and assigned to the grid.
The 15-class land use data was compiled from 30m resolution U.S. National Land Cover  Database (NLCD).

\section{Methods}
\label{sec:2}
\subsection{Gridded Surface Subsurface Hydrologic Analysis (GSSHA) Model Setup}
\label{sec:gssha}
GSSHA is developed and actively operated by the Engineer Research and Development Center (ERDC) of the United States Army Corps of Engineers (USACE).
GSSHA is an open source distributed-parameter hydrologic model capable of coupling multiple physical interactions among 1D channel flow, 2D overland flow, infiltration and groundwater flow, precipitation interception, snow melting, and evapotranspiration \cite{Downer2006}. 
Among the process-driven models, GSSHA has been widely used by many researchers for various purposes from total maximum daily loads (TMDLs) to compound flooding; in the period of 2000-2017, GSSHA has been used in more than 85 scientific/technical projects \cite{borah2018watershed}. More recently, \cite{silva2018dynamic} coupled GSSHA with the state-of-the-art surge modeling system (ADCIRC-SWAN) to simulate compound flooding on the east coast of Puerto Rico. Other recent studies used GSSHA to improve the parameterization of the Storm Water Management Model \cite{fry2018using}, and to evaluate the performance of satellite-based precipitation products in comparison to radar data \cite{furl2018assessment}.
Thus it is chosen as the benchmark model for this study.
Considering the location, geography, and objectives of the study only surface flow routing processes were activated.

In this paper, a GSSHA model was built for the study area using the Watershed Modeling System (WMS) version 10.1. WMS is a watershed RR simulation and modeling software application from Aquaveo\texttrademark \cite{daniel2011watershed}. The software supports a number of hydraulic and hydrologic models including GSSHA that can be used to create drainage basin simulations.
A uniform 2-D grid with 56,606 cells with a dimension of 100 by 100 meters for 2D overland flow was constructed. 
The streams were represented by 49 reaches of trapezoidal channels. The channel nodes have an average length of 470 meters in the longitudinal direction. The cross-section geometry is approximated based on an existing HEC-RAS model for Brays Bayou  developed by  the USACE. To compute 2D overland flow, the alternating direction explicit (ADE) method was chosen in GSSHA. To assign surface roughness parameters (Manning coefficient) an index map was created using 15-class land use data. For each land use class (see Figure \ref{fig:land_use} in the supplementary information (SI)), Manning coefficient recommended by the National Resource Conservation Service (NRCS) was used (see Table \ref{tab:land_use} in the SI). The channel flow is modeled using explicit diffusive wave method. The precipitation data was obtained from 16 rainfall gages inside the Brays Bayou watershed. The distributed rainfall was then interpolated using Theissen polygon and inverse distance weighted methods.

Limited by the computational expense, the GSSHA model used in this study was calibrated on an event from 11/17/2016 to 11/27/2016. The peak flow rate of this event was 82.69 cms, which was close to that of the moderate rainfall event. River channel's Manning's coefficient was set as the only calibration factor and restricted within the range between 0.001 and 0.02. The GSSHA built-in automated calibration tool using Levenberg Marquardt (LM) / Secant LM (SLM) was chosen as the optimization algorithm for model calibration. Optimization started with Manning's n at 0.02 and RMSE at 7.90 cms. After 12 model runs, 0.003 was found to be the optimal Manning's coefficient. The RMSE of the final model's prediction was 6.45 cms.

Furthermore, four flood events in 2017 with different scales were simulated using the calibrated model to be compared with the LSTM model's result and observed data. To be consistent with the data-driven model, discharge at the chosen USGS fresh water gage (08075000) is set as the observation point.

\subsection{Long-Short-Term-Memory (LSTM) network}
\label{sec:lstm}
\begin{figure}[h]
    \centering
    \begin{subfigure}{0.33\textwidth}
        \includegraphics{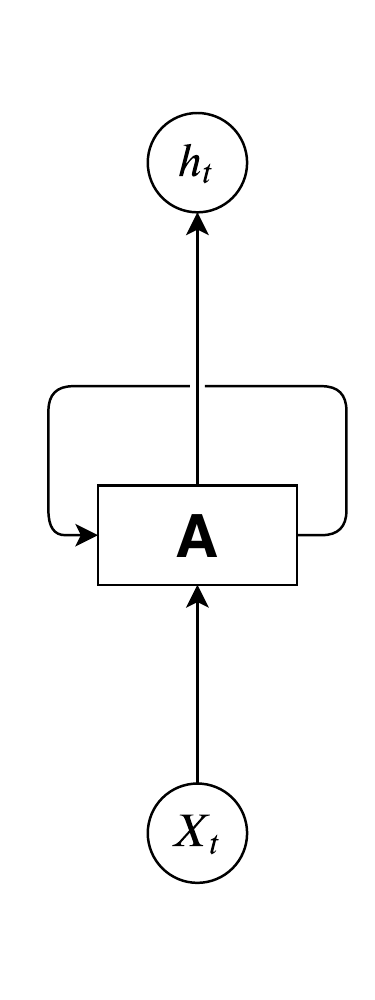}
        \caption{Structure of RNN.}
        \label{fig:rnn_rolled}
    \end{subfigure}
    \begin{subfigure}{0.66\textwidth}
        \includegraphics{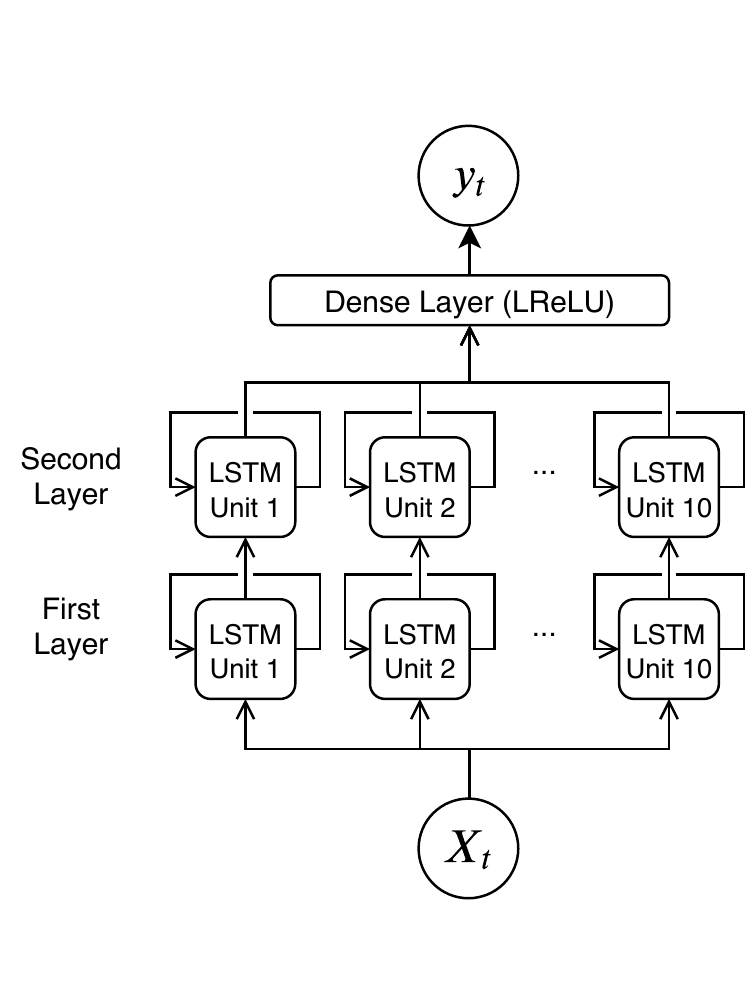}
        \caption{Developed two layer LSTM network with 10 hidden units.}
        \label{fig:lstm-network}
    \end{subfigure}
    \caption{Architecture for recurrent network and the developed LSTM network}
\end{figure}

In this study, a standard LSTM network was used to predict the discharge from rainfall data. The LSTM network is an RNN composed of LSTM units. RNN structures have been explained elsewheres \cite{lecun2015deep}, but in brief and as shown in Figure \ref{fig:rnn_rolled}, at each time step $t$, a neural network, $A$, looks at some input $x_t \in \mathbb{R}^d$, where $d$ is the dimension of the input, and hidden state from the last time step $h_{t-1}$, and outputs a value $h_t \in \mathbb{R}$. A loop allows information to be passed from one step of the network to the next. At the next time step $t+1$, the new input $x_{t+1}$ and hidden state $h_t$ are fed into the network, and new hidden state $h_{t+1}$ is computed. In theory, RNNs are capable of handling "long-term dependencies". For instance, initial input $X_0$ could affect the hidden state value 500 steps later ($h_{500}$). Unfortunately, in practice, due to numerical limitation during the optimization stage, RNNs consisting of single layer of artificial neurons are unable to learn to connect the long-term dependency \cite{bengio1994learning}. Instead of an artificial neuron, an LSTM unit contains a memory cell $g_t$ and three gates. These gates are input gate $i_t$, forget gate $f_t$, and output gate $o_t$ \cite{gers1999learning}.

At time step $t$, the LSTM unit takes input $x_t$, hidden states $h_{t-1}$. Then it updates the hidden states following (\ref{eq:lstm-unit}):
\begin{equation} \label{eq:lstm-unit}
\begin{aligned}
&i_t = \sigma (W_{ii} x_t + b_{ii} + W_{hi} h_{t-1} + b_{hi}) \\
&f_t = \sigma (W_{if} x_t + b_{if} + W_{hf} h_{t-1} + b_{hf}) \\
&g_t = \sigma (W_{ig} x_t + b_{ig} + W_{hg} h_{t-1} + b_{hg}) \\
&o_t = \sigma (W_{io} x_t + b_{io} + W_{ho} h_{t-1} + b_{ho}) \\
&c_t = f_t c_{t-1} + i_t g_t \\
&h_t = o_t \tanh (c_t) \\
\end{aligned}
\end{equation}
where $\sigma$ is the sigmoid function; $W_{ii}$, $W_{if}$, $W_{ig}$, and $W_{io}$ are the input-hidden weights; $W_{hi}$, $W_{hf}$, $W_{hg}$, and $W_{ho}$ are the hidden-hidden weights; $b_{ii}$, $b_{if}$, $b_{ig}$, and $b_{io}$ are the input-hidden biases; $b_{hi}$, $b_{hf}$, $b_{hg}$, and $b_{ho}$ are the hidden-hidden biases.

Depending on the size of the watershed, the peak of generated runoff can be observed from a couple of hours to a couple of weeks after the event. For instance, for a time step of 15 minutes, a rainfall event that lasted for a week would have more than 650 steps in time. To accurately model the RR process, the model is required to memorize the effect of precipitation from the beginning of the event, which is numerically difficult for standard RNN. Thus, Such a structure shown in (\ref{eq:lstm-unit}) is proposed to cope with the vanishing gradient problems that can be encountered when training standard RNNs \cite{gers1999learning}. With the objective of channel discharge prediction, a two-layer LSTM network with 10 hidden units in each layer was designed as shown in Fig \ref{fig:lstm-network}.
A standard LSTM implementation from the open source deep learning platform PyTorch \cite{Paszke2017} was adopted in this study to develop the data-driven model.

In this study, input $x_t = (x_{t_1}, x_{t_2}, ..., x_{t_n})$ is the vector of precipitation gage readings at time $t$. $x_{t_i}$ corresponds to the reading of the $i$-th precipitation gage at time $t$. 
Note that flow at any of the previous time steps ($y_{t-k}$) is not within the input vector indicating the developed model is not a time series model. In other words, it does not depend on the immediate past observation to make prediction. 
For each time step $t$, the precipitation readings are input into the network, and $h_t \in \mathbb{R}^p$ is computed by the two layers of LSTM network $\mathcal{M}_{LSTM}$: $h_t, c_t = \mathcal{M}_{LSTM} (x_t, h_{t-1}, c_{t-1})$. Each hidden unit works independently. Formally, the LSTM network $\mathcal{M}_{LSTM}: \mathbb{R}^{2p} \mapsto \mathbb{R}^{2p}$ is defined as:
\begin{equation} \label{eq:lstm-network}
\begin{aligned}
&h_{l_t} = o_{l_t} \tanh (c_{l_t}) \\
&c_{l_t} = f_{l_t} c_{l_{t-1}} + i_{l_{t}} g_{l_{t}}
\end{aligned}
\end{equation}
where dimension of $h_t$ and $c_t$, $p=10$ is the number of hidden units set by user. $h_{l_t}, c_{l_t}$ are the $l$th element of the $h_t, c_t$ vector. The gates and cell state $o_{l_t}, f_{l_t}, i_{l_t}, g_{l_t}$ are updated following (\ref{eq:lstm-unit}). Since the output dimension is 1 (we are only trying to predict the outlet discharge), another fully connected layer $\mathcal{M}_{fc}:\mathbb{R}^p \mapsto \mathbb{R}$ would transform the $h_t$ to the output $\hat{y_t}$: $\hat{y_t} = \mathcal{M}_{fc} (h_t)$:
\begin{equation}
    \mathcal{M}_{fc} (h_t) = g_a (\beta_0 + \sum_{l=1}^p \beta_l h_{l_t})
\end{equation}
where $g_a$ is the activation function; in the context of artificial neural networks, the activation function of a node defines the output of that node for a given input or set of inputs. A nonlinear activation function allows the neural networks to model nonlinear relationship between the input and output. This function is also known as the transfer function \cite{lecun2015deep}. An activation function defined as the positive part of its argument is called rectifier:
\begin{equation}
    g_a (x) = x^+ = max(0, x),
\end{equation}
where x is the input to a neuron. Leaky rectified linear unit (LReLU) is a generalization of rectifier where a small, positive gradient is allowed when the unit is not \textit{active} (input to the neuron is not positive):
\begin{equation}
    g_a (x) = \begin{cases}
x \ \text{if} \ x > 0, \\
0.01x \ \text{otherwise}.
\end{cases}
\end{equation}

As shown in (\ref{eq:lstm-unit}) and (\ref{eq:lstm-network}), an initial hidden state is required for the LSTM network to start a forward propagation, i.e. $h_1, c_1 = \mathcal{M}_{LSTM} (x_1, h_0, c_0)$. After some preliminary test runs, a modification was made to the model to set the initial hidden states $(h_0, c_0)$ as learnable parameters instead of randomly initializing it to improve the prediction performance at the initial stage. The initial hidden states $(h_0, c_0)$ are fixed after training.
This change was justified because hydrologic models are generally set up with an initial condition (base flow) which exists before the rainfall event. 

Like any supervised learning algorithm, calibrating the LSTM model requires \textit{training} the model by optimizing the objective function $\mathcal{L} (\hat{y_t}, y; w)$ under some constraint $C$, where $\hat{y_t}$ and $y$ are the prediction and observation respectively. $w$ is the learnable parameter:

\begin{equation}
    \begin{aligned}
    & \underset{w}{\text{arg min}}
    & & \mathcal{L} (\hat{y_t}, y; w) \\
    & \text{subject to}
    & & w \in C.
    \end{aligned}
\end{equation}

In this study, the objective function is defined as the mean square error (MSE) between the prediction and observation:

\begin{equation}
    \mathcal{L} (\hat{y_t}, y; w) = \frac{1}{n} \sum_{t=1}^n (\hat{y_t} - y)^2.
\end{equation}

To avoid over-fitting of the deep learning models, regularization by adding penalty to the learnable parameters is a widely used approach   \cite{lecun2015deep}. In practice, regularization $\mathcal{R}(w)$ is added to the objective function. A regularized version of the optimization problem becomes,

\begin{equation}
\begin{aligned}
& \underset{w}{\text{arg min}}
& & \mathcal{L} (\hat{y_t}, y; w) + \mathcal{R}(w) \\
& \text{subject to}
& & w \in C.
\end{aligned}
\end{equation}
where $C$ is the constraint on $w$.

In this work, $l_2$ regularization is used to prevent the model from over-fitting to the training data,
\begin{equation}
    \mathcal{R}(w) = \lambda ||w||_2^2 / 2.
\end{equation}
where  $\lambda$ is the regularization parameter ($10^{-6}$). Larger $\lambda$ corresponds to more regularization. Note that this technique is also known as weight decay because when applying standard stochastic gradient descent (SGD), it is equivalent to updating the weight in this way:
\begin{equation}
    w_{i+1} = w_i - \lambda w_i - \alpha \frac{\delta L}{\delta w}|_{w_i}
\end{equation}
where $w_i$ is the learnable parameters at step $i$, $\alpha$ is the learning rate, and $\frac{\delta L}{\delta w}|_{w_i}$ is the stochastic gradient approximation at step $i$. Thus, at each step, the weight $w$ \textit{decays} by $(1-\lambda)$. For standard stochastic gradient descent (SGD), weight decay can be made equivalent to $l_2$ by a reparameterization of the weight decay factor based on the learning rate. However, this is not the case for adaptive gradient descent methods including Adam optimization \cite{kingma2014adam}, which is used in this study with a learning rate of $10^{-4}$.

\subsection{LSTM Model Training, Validation, and Evaluation}
In machine learning, a mathematical model is built from existing data. However, the task of the machine learning model is to make predictions on future data that is not available at the model construction time. To evaluate the model performance on unseen data, a common practice in supervised machine learning is to split the data into three data sets which are used in different stages of the creation of the model.

Specifically, the model is initially fit on a training dataset, that is a set of examples used to fit the parameters (e.g. weights of connections between neurons of artificial neural networks) of the model. Successively, the fitted model is used to predict responses for the observations in a second dataset called the validation dataset (e.g. predict hydrograph given precipitation in this study). The validation dataset provides an unbiased evaluation of the model fit on the training dataset while tuning the model's hyperparameters (e.g. the number of the hidden units in neural network, number of LSTM layers, regularization, etc.). The combination of the hyperparameters with the best validation performance is then chosen for the machine learning model. Finally, the test dataset is used to provide an unbiased evaluation of a final model. If the examples from the test dataset have never been revealed to the model during training and validation stages, the test dataset is also called a holdout dataset.

\begin{figure}[h]
    \centering
    \includegraphics[width=\textwidth]{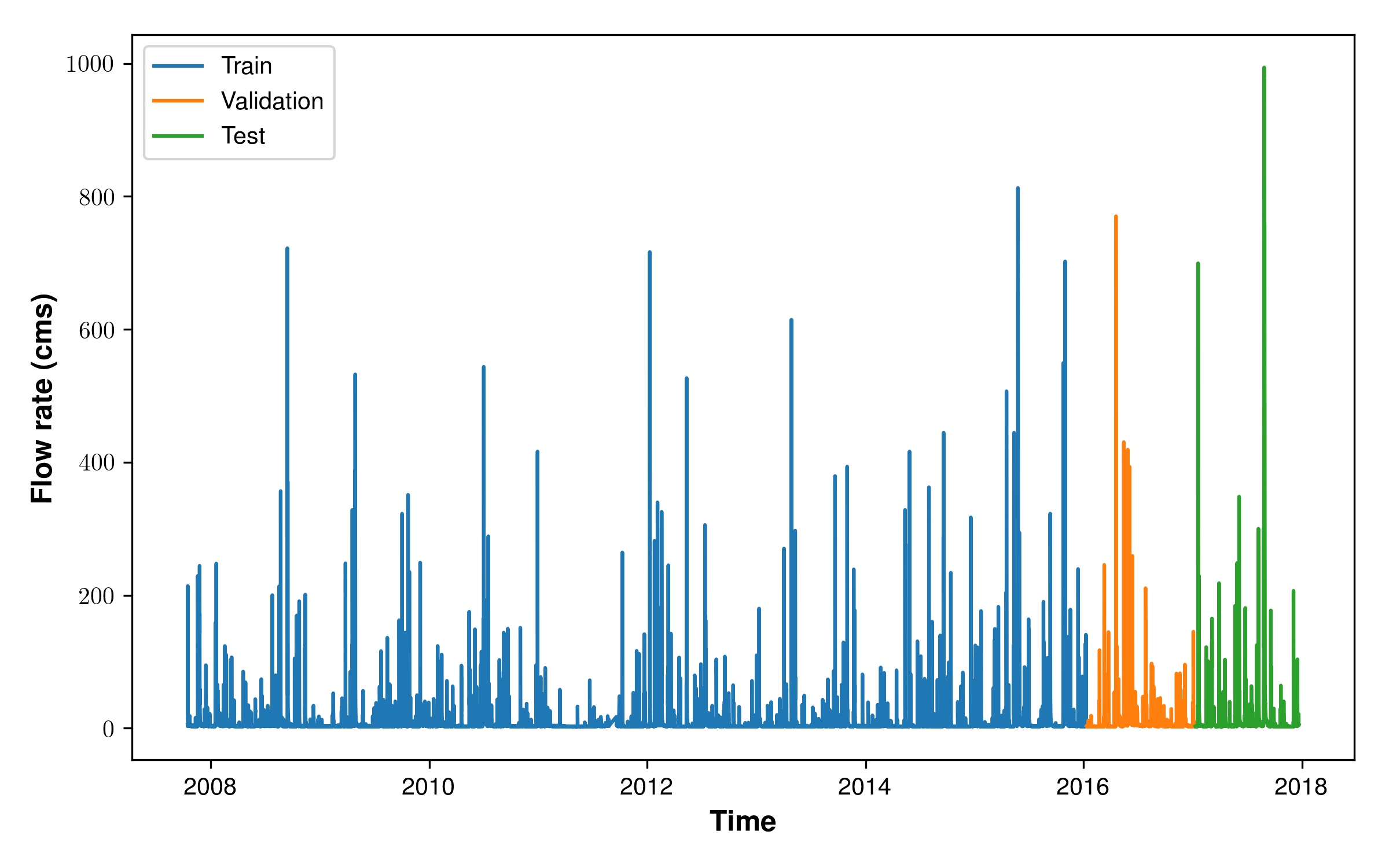}
    \caption{Training, validation, and test split.}
    \label{fig:train-test-split}
\end{figure}

Hydrologic data was split into train, validation, and test data sets. As shown in Figure \ref{fig:train-test-split} all 15-minute data up to the end of 2015 (2007-2015) was used for training. The entire year 2016 was used for validation and 2017 was used as the holdout test dataset. This train-validation-test split scheme is designed to minimize over-fitting and consistent with realistic prediction scenarios.

Particularly at the training stage, each time series is considered as a training example. A training \textit{iteration} includes a forward propagation of the training example that computes the output, a backward propagation that computes the gradient, and an optimization step that updates learned parameters. Multiple training examples can be put into a \textit{batch} where both forward and backward propagation are processed in parallel respectively. One forward pass and one backward pass of all the training examples is called an \textit{epoch}. More details can be found in \cite{lecun2015deep}.

During training, each sequence of data from the training dataset was input into the LSTM network as one batch. Built-in Adam optimization algorithm was used to optimize the MSE loss function. To avoid over-fitting, an $l_2$ regularization with coefficient of $10^{-6}$ was added for all learnable parameters.  The learning rate  was  set  to $10^{-4}$. These hyper-parameters were used for all tests for this study unless otherwise indicated.

To handle the missing data, a threshold of 90-minutes was set. If the missing data gap was less than or equal to 90 minutes (6 missing points), the missing points were imputed by linear interpolation. For gaps greater than 90 minutes, the sequence was split at the gaps. To speed up training and avoid the gradient exploding problem \cite{sutskever2014sequence}, the time series is further split into even shorter series.

To help the data-driven model extrapolate better, a minimum-maximum scaling was applied to both input and output variables \cite{hsu1995artificial,hettiarachchi2005extrapolation} so that all transformed variables are in the range of [0, 0.9]. 
Preliminary tests showed the model trained by the transformed variables in the range of [0, 0.9] performed better in comparison to the one trained by variables in the range of [0, 1].

\subsection{LSTM Scenarios}

\subsubsection{Scenario 1: Physical Consistency of LSTM Model}

Physically, it is obvious that the amount and pattern of rainfall collected by gages near/upstream of the flow gage are more relevant to runoff discharge than those collected by gages that are far from/downstream of the river gage. 
To explore how the LSTM model results are spatially distributed and verify that the trained data-driven model is consistent with the physical intuition, two numerical tests were conducted using precipitation data from all 153 aforementioned rainfall gages:
\begin{enumerate}
    \item Precipitation data from each rainfall gage was used to train a separate LSTM model. Then the training loss of all of the 153 models were recorded and compared. The assumption was: if the LSTM model could actually learn the physical correlation between precipitation and river discharge, the model trained with more relevant input data should perform better.
    
    \item All gages' data were used to train a single LSTM model. It is natural to assume that a physically consistent model should pay more attention to the more \textit{important} gages.
\end{enumerate}

The purpose of these two numerical tests is to explore the characteristics of the LSTM network for the defined application and reduce the number of input gages to the model based on the physical intuition of the problem.

For the first test, each model was trained for 200 epochs on the training data set combined with the validation data set and the best performing epoch with the minimum training error was recorded. Because the models trained in this test was not intended to be used for prediction, regularization and validation were not applied in this case. 
For the second test, due to the higher input dimension more epochs were required for the model to converge. Thus, the all-gage model was trained for 400 epochs on the training data set. The best performing epoch on the validation data set was chosen as the trained model.  Regularization and validation were applied here to (1) cope with the ill-conditioning problem when highly correlated precipitation data from different gages are presented; (2) prevent over-fitting so that the model parameters including the first layer weights, come from a meaningful model. As noted before, the first layer of the LSTM network takes precipitation input. After the training, all the learnable input-hidden weights ($W_{ii}, W_{if}, W_{ig}$, and $W_{io}$) of the first LSTM layer are grouped by gages and then flattened to a vector W, i.e.
\begin{equation}
    W = [W_{ii,1}, W_{if,1}, W_{ig,1}, W_{io,1}, ..., W_{ii,n}, W_{if,n}, W_{ig,n}, W_{io,n}]
\end{equation}
where $n$ is the number of hidden units, i.e. 10. For each gage, three parameters of the learnable input-hidden weights were defined by the $l_1, l_2$, and $l_{\infty}$ norms of $W$:

\begin{equation}
        ||W||_1 = \sum_{i=0}^{n} (|W_{ii}| + |W_{if}| + |W_{ig}| + |W_{io}|)
\end{equation}
    
\begin{equation}
    ||W||_2 = [\sum_{i=0}^{n} (W_{ii}^2 + W_{if}^2 + W_{ig}^2 + W_{io}^2)]^{1/2}
\end{equation}

\begin{equation}
\begin{aligned}
    ||W||_{\infty} = \max \{&|W_{ii,1}|, |W_{if,1}|, |W_{ig,1}|, |W_{io,1}|, ..., \\
    &|W_{ii,n}|, |W_{if,n}|, |W_{ig,n}|, |W_{io,n}|\}
\end{aligned}
\end{equation}

Thus, for each gage, this test generates four parameters: the training error $e$, and the three norms of the weight vector $W$. To find any correlation, if any, among the three norms of the weight vector and training errors from the first numerical test, a correlation analysis was conducted using both Pearson correlation coefficient ($r$) and Spearman’s rank correlation coefficient ($\rho$).

\subsubsection{Scenario 2: LSTM model using 10 rainfall gages}

To reduce the training time and the need for input data, it is necessary to reduce the number of gages used for the training. In addition, reducing the number of gages should not decrease, if not increase, the performance of the model. An exhaustive feature selection would require trying all combinations of gages which means training $2^{153}$ models which is infeasible. Thus, the choice of rainfall gage was the 10 most relevant gages determined by scenario 1 using the gages with the minimum training errors.

A slightly different training process for this test was followed since this model was intended to be used for prediction. The LSTM model was trained on the training data set and regularization was added. The number of  epochs were restricted to 200 and the best performing (in terms of evaluation score) epoch on the validation data set was chosen as the training result.

To show the causal improvement of this feature selection approach, more numerical tests were conducted. Comparison was made among models trained with the 10 best gages (based on training error), 10 randomly sampled gages (sampled 5 times) from all 153 gages, 10 randomly sampled gages within the watershed, and 10 closest gages to the discharge gage. Thus, a total of eight models using 10 rainfall gages were built and tested.

\subsection{Analysis and comparison of LSTM and GSSHA models}
Since this study focuses on RR prediction for flood events, the evaluation was focused on flood events instead of normal flow regime dominated by the base flow and tidal mechanisms. The Nash–Sutcliffe model efficiency (NSE) and root-mean-square error (RMSE) were selected as the metrics for model evaluation. As introduced in Section \ref{sec:lstm}, our developed model (as well as GSSHA model) is not a time series forecasting model. Hence the metrics do not include those evaluations for time series models such as persistency criterion which compares the forecasting with the predictions from a naive persistence model.

To compare the LSTM model with the benchmark GSSHA model, four events of different scale in terms of precipitation and river discharge were chosen from the test data set (2017); low rainfall event  from 9/28/2017 to 10/6/2017, moderate rainfall event from 12/16/2017 to 12/19/2017, high rainfall event from 12/2/2017 to 12/12/2017, and finally, an extreme rainfall event including Hurricane Harvey, which started from August 23 and ended on September 1, 2017.
Note that the moderate event follows the high event with an interval of 4 days. Here it was assumed the precipitation of the first event had completed runoff by the time the second event starts.
Again, the choice of these four events were due to the limitation of computational expense of the GSSHA model, not the LSTM model. The LSTM model has no such restriction and was tested for all of 2017's flood events.

\section{Result and Discussion}

The results and discussion section will start by presenting the results from the developed LSTM model, how well the model can predict runoff discharge using precipitation from each gage, and how consistent the results were with physical intuition. For this purpose the  two numerical tests of experiment 1 will be discussed. Next, the performance of the constructed LSTM model with the 10 selected gages' data as input will be investigated and the results will be compared with the model using all 153 gages' data as input. 
Finally, the results of the GSSHA model developed for the Brays Bayou watershed will be presented and the difference between the predictions from GSSHA and LSTM will be discussed.

\subsection{Physical consistency of LSTM result}

\begin{figure}[h]
    \centering
    \includegraphics[width=0.8\textwidth]{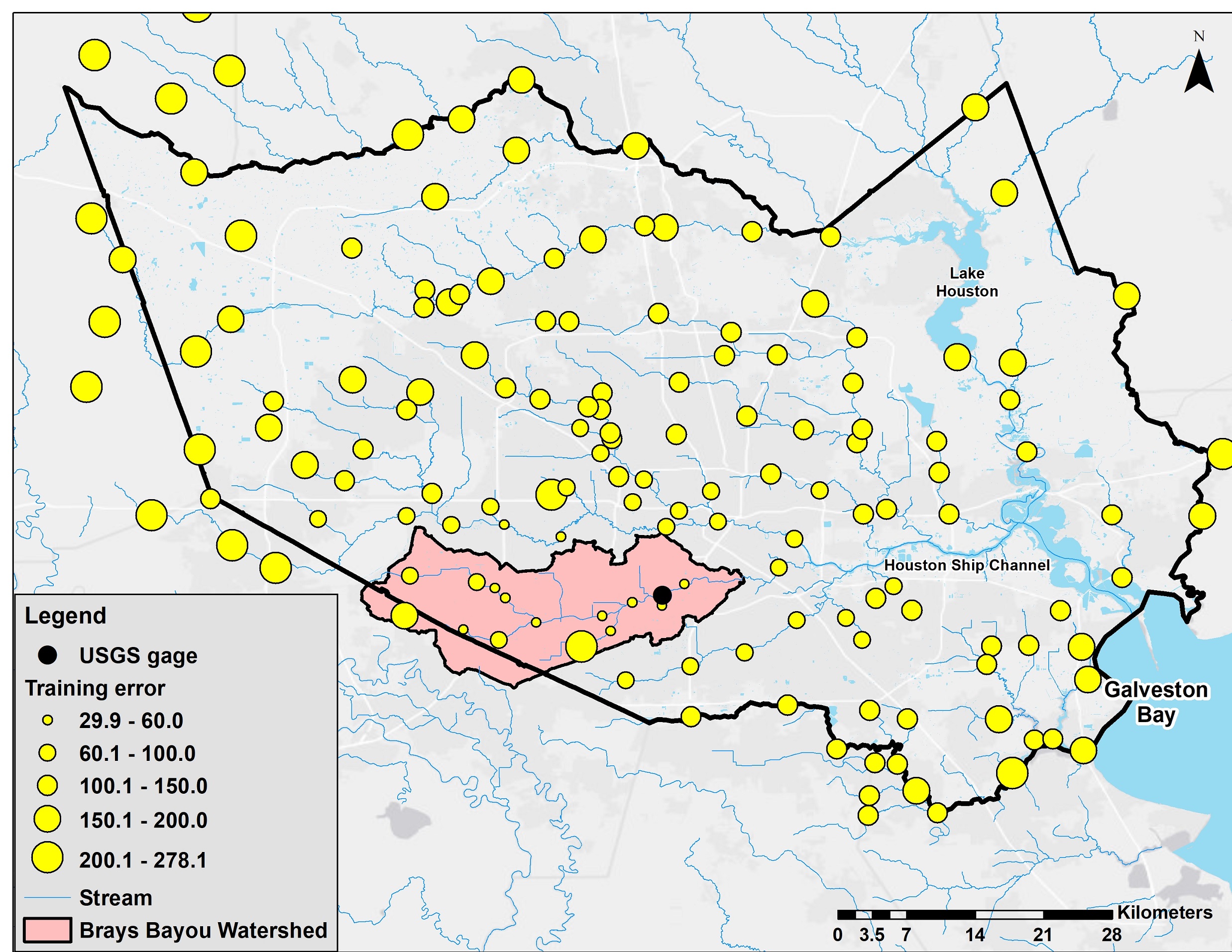}
    \caption{Training error map for single gage models}
    \label{fig:training_err}
\end{figure}

The training errors of LSTM models using each single rainfall gage (first numerical test in experiment 1) are shown in Figure \ref{fig:training_err}. The lowest training error was 29.94 in a gage just upstream of the discharge gage and highest training error was 278.11 in a gage located outside of Harris County. From Figure \ref{fig:training_err} it can be seen that gages with the best performance are the ones located within or near the watershed. In fact, the Pearson correlation between the training error extracted from the LSTM model and the physical distance between the rainfall gages and the USGS gage was significant with a p-value of 3.5E-31 and $r$=0.77. This results show that similar to the process-driven model, where precipitation drives runoff and the amount of precipitation falls into the watershed is represented by the interpolation of the rainfall gage recording, the data-driven model also performs better when better representation of the distributed precipitation is provided.

The correlation between training error $e$ and the three parameters in scenario 1 could provide intuition on how much attention the LSTM model pays to the important rainfall gages. This makes more sense when considering the fact that the spatial distribution of the \textit{best performing gages} matches very well with the physical intuition. The Pearson correlation coefficient, Spearman's rank correlation coefficient, and the respective p-values of the statistical t-tests are shown in Table \ref{tab:corr}. All statistical t-tests suggest it is safe to reject the null hypothesis that the weight parameter is uncorrelated with the gage training error. Moreover, there is a non-trivial negative correlation between the norms of first layer weights and performance of model trained using the corresponding gage. The result suggests that statistically the LSTM model pays more attention to the \textit{physically important} gages than those irrelevant gages.

\begin{table}[h]
    \caption{Correlation between gage training error and weights parameters.}
    \centering
    \begin{tabular}{l c c c c}
    \hline
    Parameter    & $r$    & p-value of $r$ & $\rho$ &  p-value of $\rho$ \\
    \hline
    $||W||_1$   & -0.731 & 7.551e-27 & -0.335 & 2.264e-5 \\
    $||W||_2$   & -0.742 & 4.457e-28 & -0.343 & 1.389e-5 \\
    $||W||_{\infty}$ &  -0.664 & 9.071e-21 & -0.318 & 6.097e-5 \\
    \hline
    \end{tabular}
    \label{tab:corr}
\end{table}

As suggested in \cite{lindstrom1997development,anctil2006improvement}, eliminating redundant gages effectively improves the predictions of RR models. 
The consistency not only suggests the LSTM model is paying more attention to the more important gages, but also provides an efficient way of choosing rainfall gages for the LSTM model. Unlike the dedicated studies of choosing rainfall gages for RR modeling using areal rainfall optimization \cite{anctil2006improvement}, the LSTM model can provide a coarse yet fast approach to pick the most relevant gages.

\subsection{LSTM prediction}

\begin{figure}[h]
    \centering
    \includegraphics[width=0.95\textwidth]{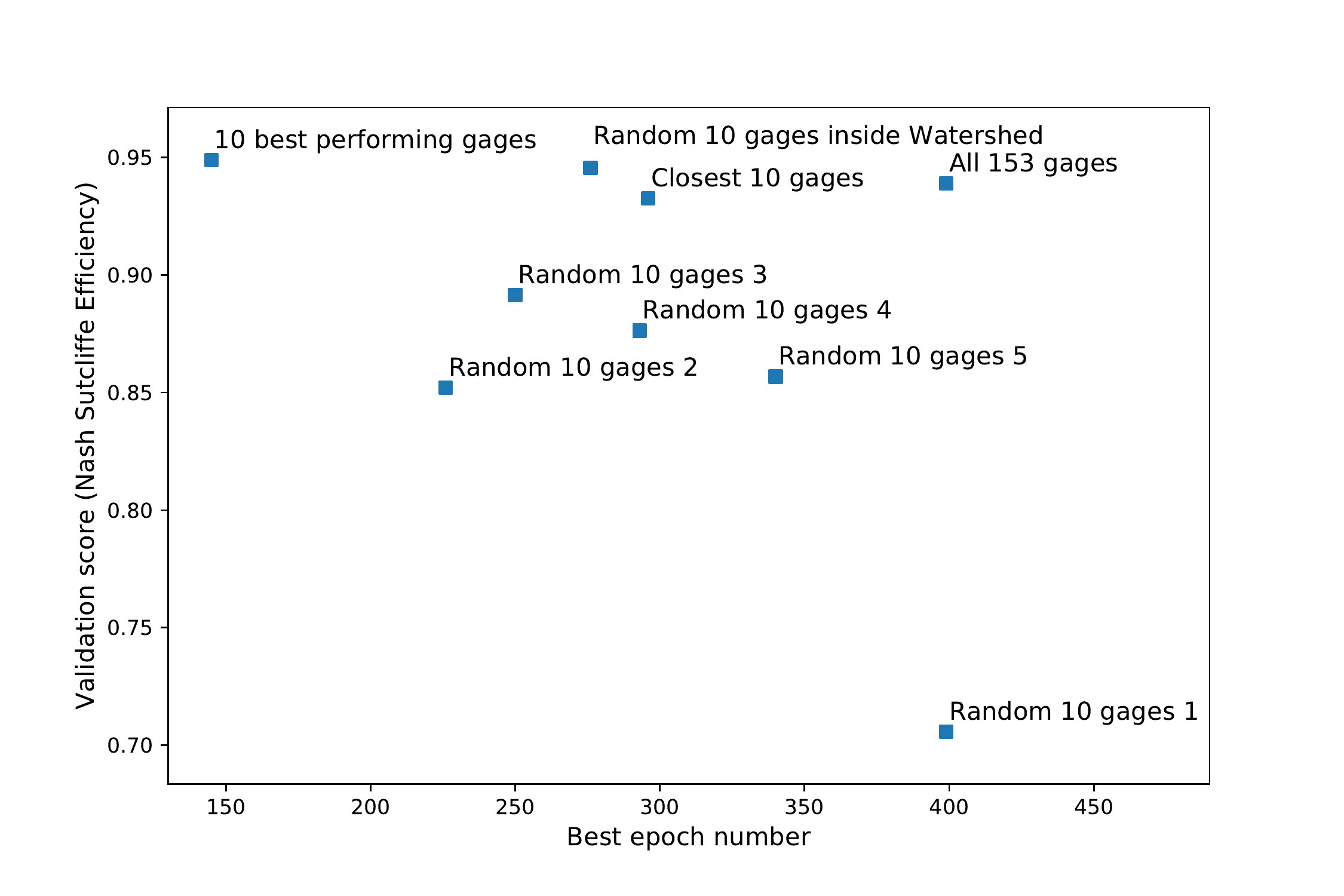}
    \caption{Validation score vs number of epochs.}
    \label{fig:feature-selection}
\end{figure}

As noted before, an exhaustive feature selection among the rainfall gages is infeasible. Using the feature selection criterion suggested in scenario 1, the 10 gages with the lowest training error in the first test, which are mostly located inside the watershed (the only gage outside the watershed is also very close to the watershed boundary), were picked as the input of the finalized LSTM model. 
To show the causal improvement of feature selection, we also compared the model performance (see Figure \ref{fig:feature-selection}) with models using (1) 10 randomly chosen gages within Harris County (sampled 5 times) (2) 10 randomly chosen gages within the watershed (3) closest 10 gages. The result in Figure \ref{fig:feature-selection} shows that the best 10 gages model not only has the best validation score, but also converged faster than all other aforementioned models. Note that the model with 10 randomly chosen gages within the watershed also has a high validation score (0.945) but this model shares 6 common gages with the best 10 gages model. Nonetheless, its best epoch is 296 which is more than twice as much as that of the best 10 gages model indicating longer computational time for training.

Compared to the 153-gage model in experiment 1, training of the best 10 gages model converged significantly faster: it took 145 epochs to converge whereas the 153-gage model required more than 350 epochs to converge using the same learning rate. The convergence analysis of Adam optimization algorithm is out of the scope of this paper, however, it is clear that the convergence of Adam algorithms is dimension dependent \cite{kingma2014adam}. Besides, lower dimension implies lower computational cost for each iteration. Hence, training models with fewer inputs would be more efficient.

The evaluation scores (RMSE and NSE) were computed for both 10-gages model and 153-gages model on training/validation/test data set and are shown in Table \ref{tab:lstm-score}. The 153-gages model had lower training error but higher validation/test error compared to the 10 gages model which implies more over-fitting of the 153-gages model.

\begin{table}[h]
    \caption{Evaluation scores of 10-gages model versus 153-gages model}
    \centering
    \begin{tabular}{l c c c c}
    \hline
    \multirow{2}{*}{Data set}    & \multicolumn{2}{c}{10-gages model} & \multicolumn{2}{c}{153-gages model} \\
     & RMSE & NSE & RMSE & NSE \\
    \hline
    Training & 8.11 & 0.921 & 5.65 & 0.961\\
    Validation & 7.83 & 0.947 & 8.49 & 0.938\\
    Test & 17.62 & 0.942 & 42.24 & 0.666\\
    Test excluding Hurricane Harvey & 8.32 & 0.906 & 10.04 & 0.864 \\
    \hline
    \end{tabular}
    \label{tab:lstm-score}
\end{table}

\begin{figure}[h!]
    \centering
    \begin{subfigure}{0.49\textwidth}
        \includegraphics[width=\textwidth]{{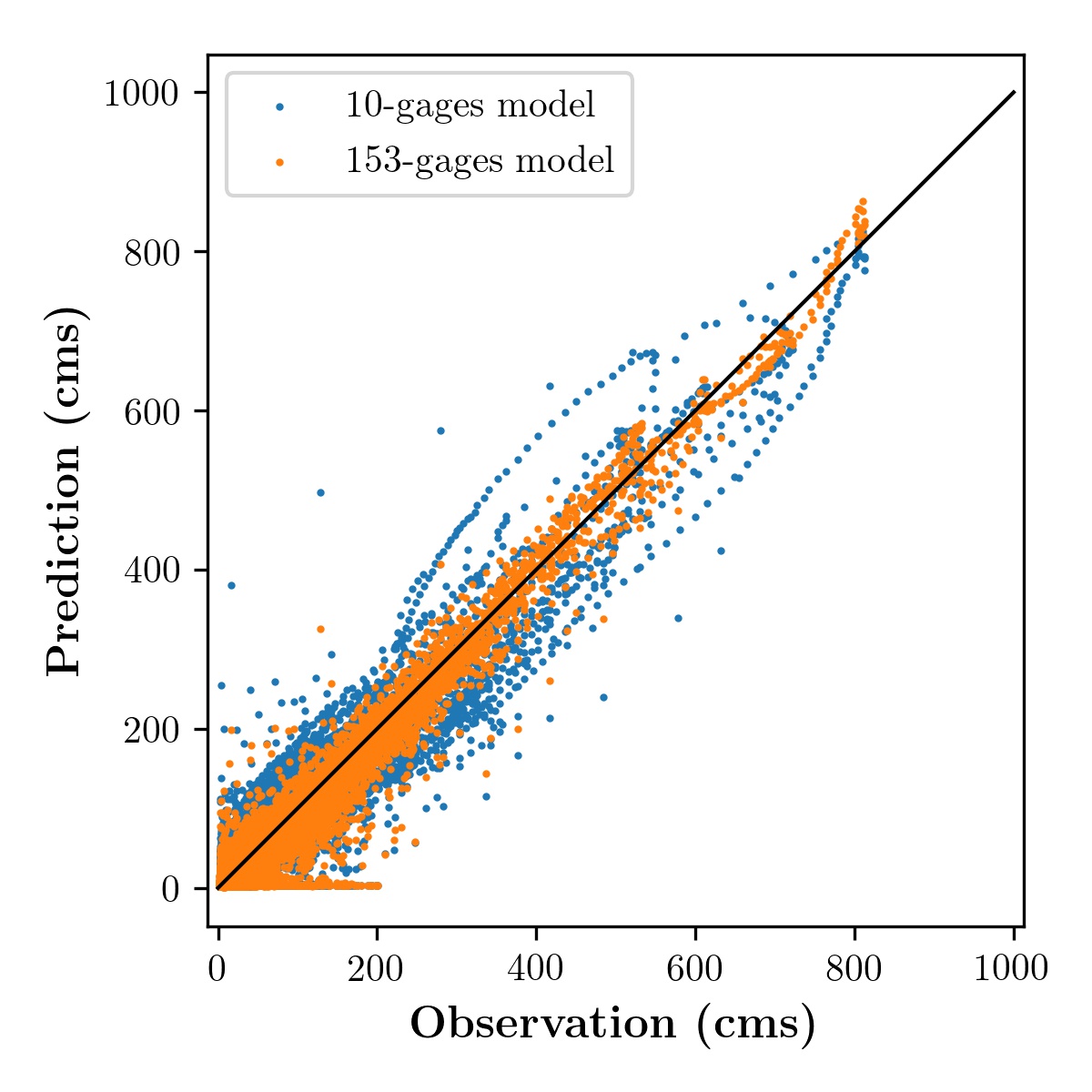}}
        \caption{Training set.}
        \label{fig:train}
     \end{subfigure}
     \begin{subfigure}{0.49\textwidth}
        \includegraphics[width=\textwidth]{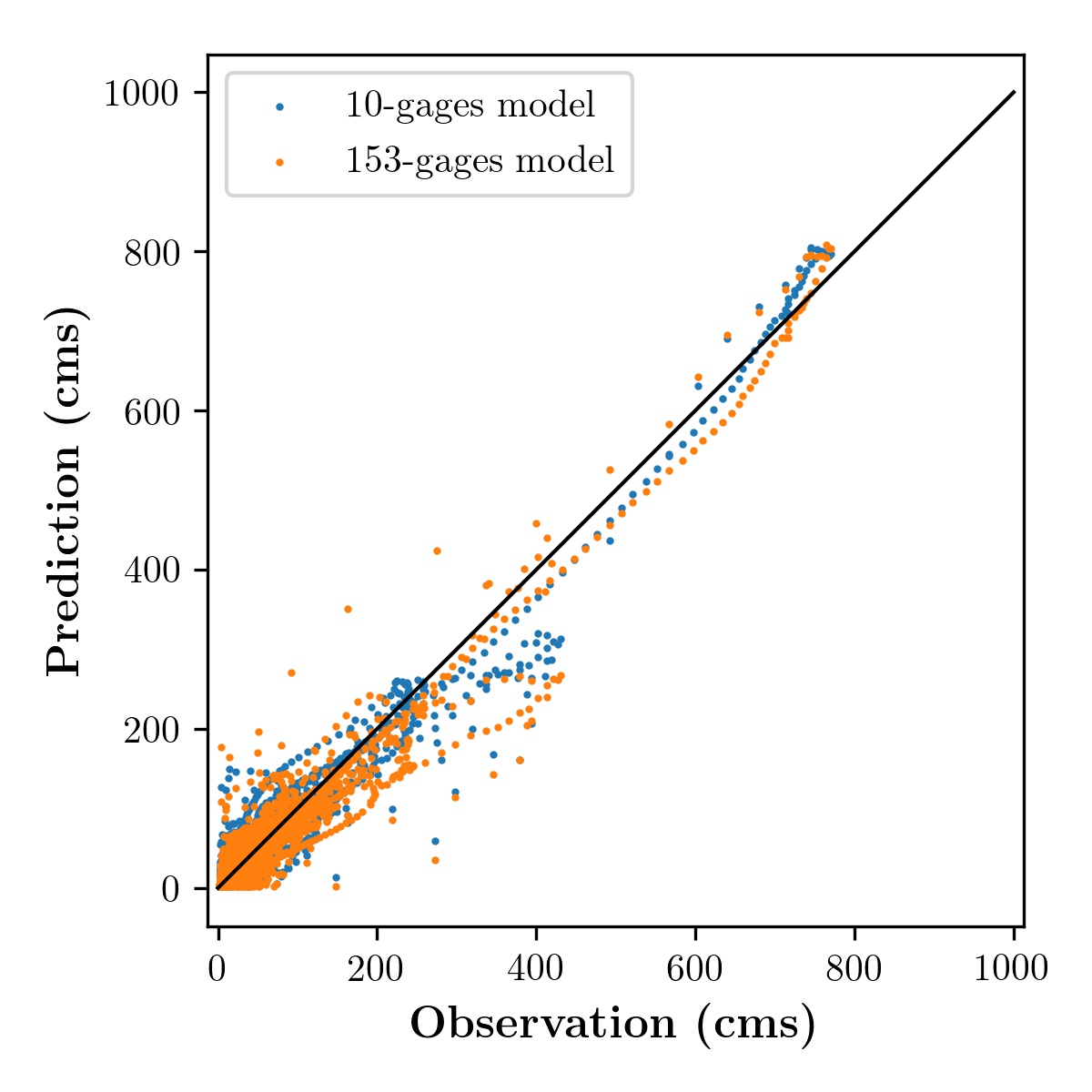}
        \caption{Validation set.}
        \label{fig:dev}
     \end{subfigure}
     \begin{subfigure}{0.7\textwidth}
        \includegraphics[width=\textwidth]{{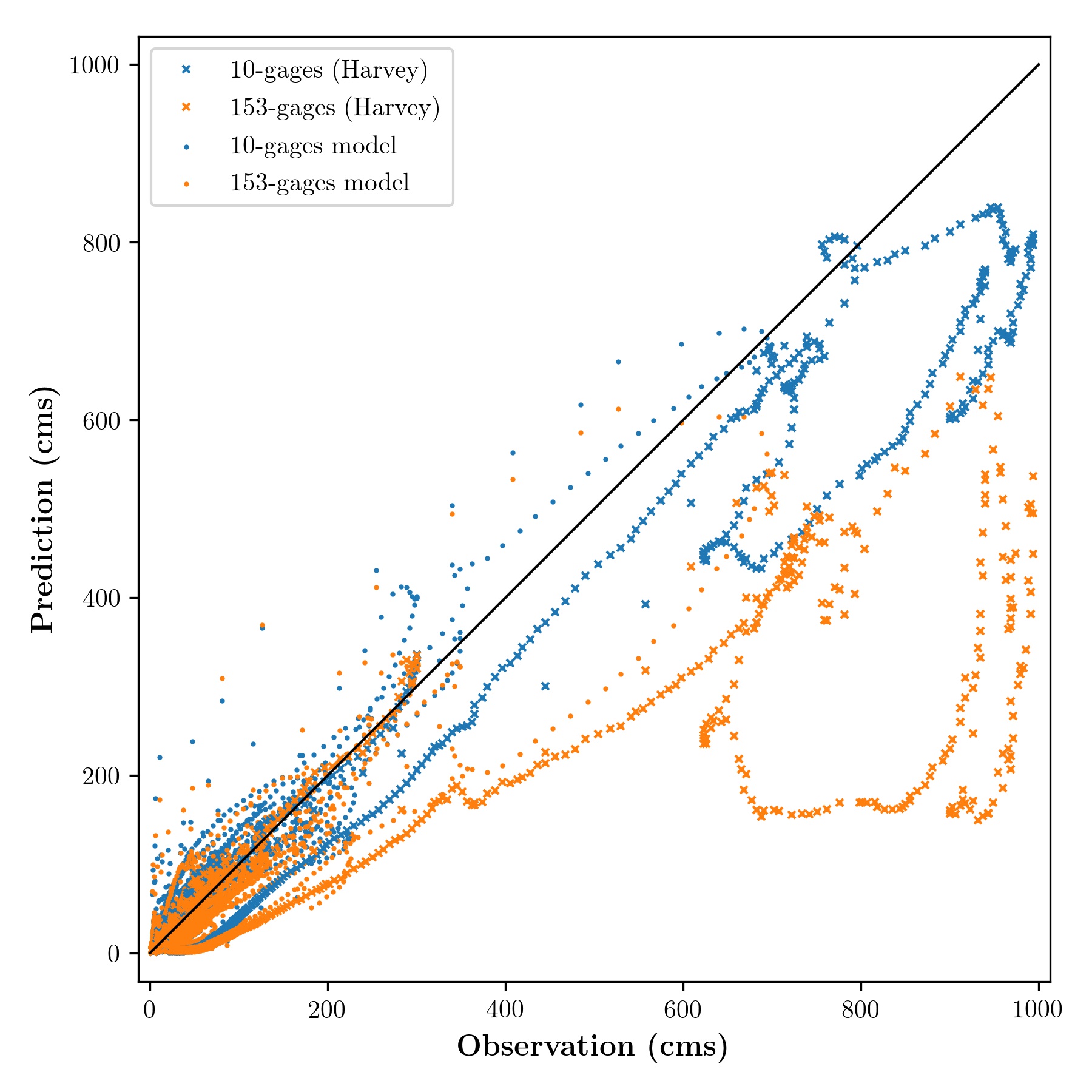}}
        \caption{Test set. Note that cross shaped dots are data points from Hurricane Harvey.}
        \label{fig:test}
     \end{subfigure}
    \caption{Scatter plot of prediction versus observation for 10-gages and 153-gages models.}
    \label{fig:scatter}
\end{figure}

Moreover, the test error of both models were significantly higher than the training and validation errors. However, this behavior is explainable and does not indicate our model is over-fitting. The larger test error was dominated by under-predicting Hurricane Harvey which was included in the test set (see Figure \ref{fig:test}). 
It should be noted that, Hurricane Harvey was an extraordinary flooding event in which, due to the high volume of precipitation, inter-basin transfer happened in many of the watersheds in the Greater Houston Area. Such phenomena is almost impossible to capture even with process-based models when only one watershed is modeled. Considering the uniqueness and rarity of Hurricane Harvey and its different behavior in both precipitation pattern and volume, the prediction of the 10-gages model on Hurricane Harvey, as shown in Figure \ref{fig:harvey}, is acceptable. The test result actually shows the relatively good extrapolation ability of the data-driven model.

From Figure \ref{fig:scatter}, it can be seen that the test set contains target discharge above 900 cms, while the training/validation sets have lower peak flow rate. The 10-gages model clearly performs better than the 153-gages model on an extreme event (Harvey). 
The flow rate versus time plots of both models for every event with peak flow larger than 30 cms are shown in the Figure \ref{fig:lstm-gage-test-all} in the SI. It can be shown that for the majority of the events except Harvey, both models are making reasonable predictions.
Table \ref{tab:lstm-score} shows that if the time series containing Hurricane Harvey was excluded from the test set, the performance of the 10-gages model would be closer to those on the training and validation set. However, compared to the 10-gages model, the 153-gages model still seems to have larger variance given that it has better training score (compared to 10-gages model) but worse validation/test score.

\subsection{Comparison of calibrated GSSHA and LSTM result} \label{sec:compare-gssha-lstm}

\begin{figure}[h!]
    \centering
    \begin{subfigure}{0.49\textwidth}
        \includegraphics[width=\textwidth]{{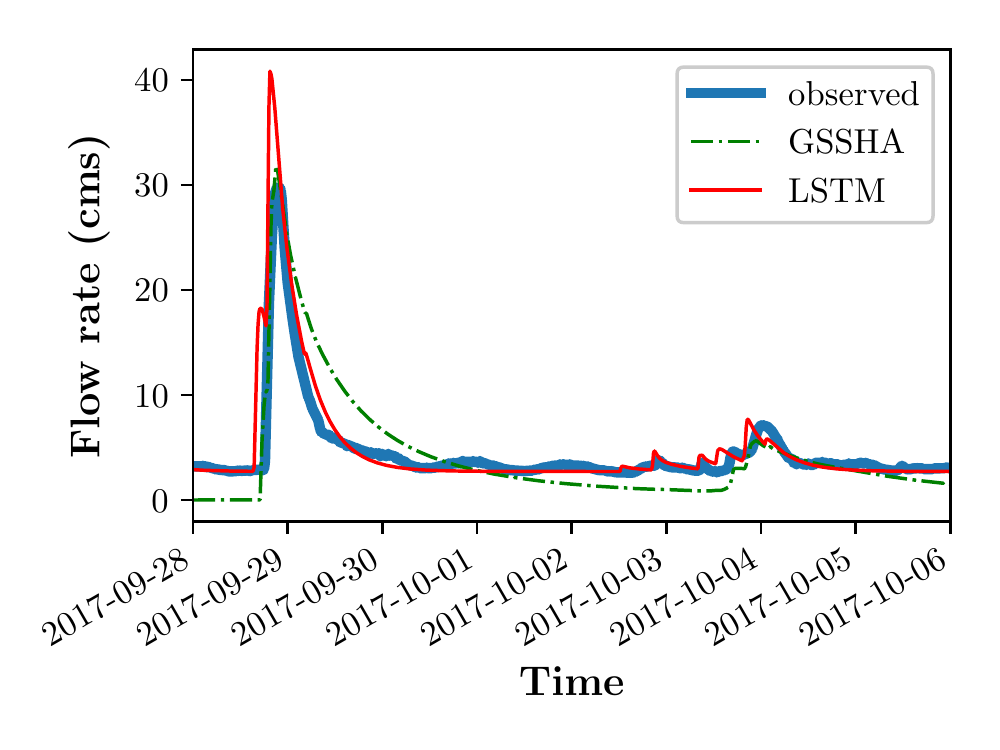}}
        \caption{Low rainfall event}
        \label{fig:very_low}
     \end{subfigure}
     \begin{subfigure}{0.49\textwidth}
        \includegraphics[width=\textwidth]{{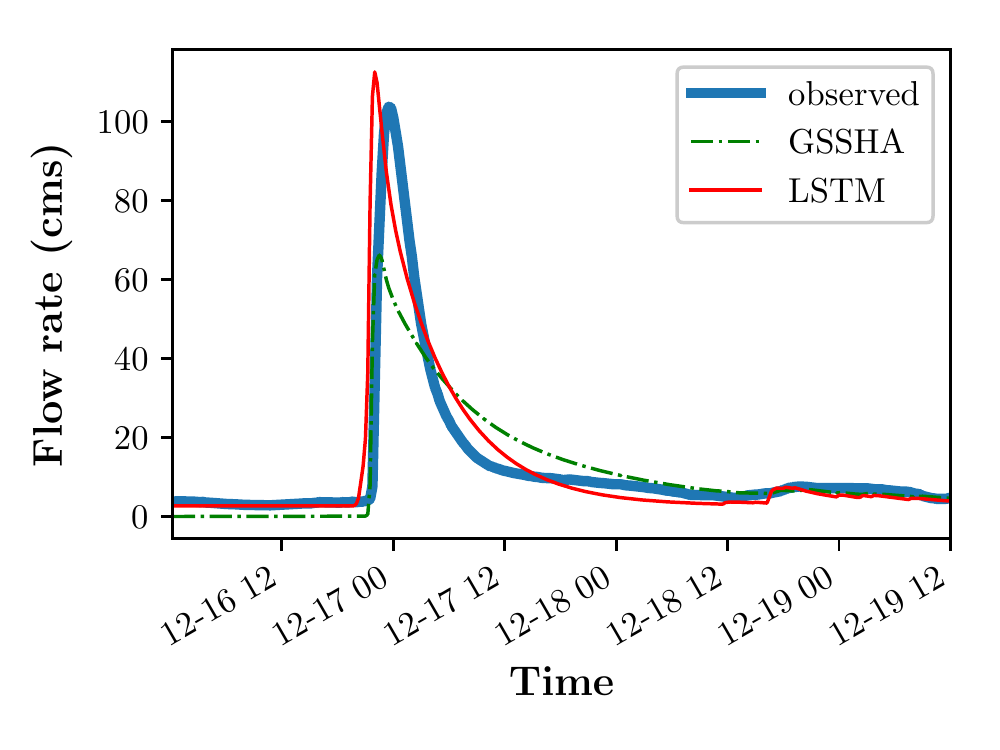}}
        \caption{Moderate rainfall event}
        \label{fig:low}
     \end{subfigure}
     \begin{subfigure}{0.49\textwidth}
        \includegraphics[width=\textwidth]{{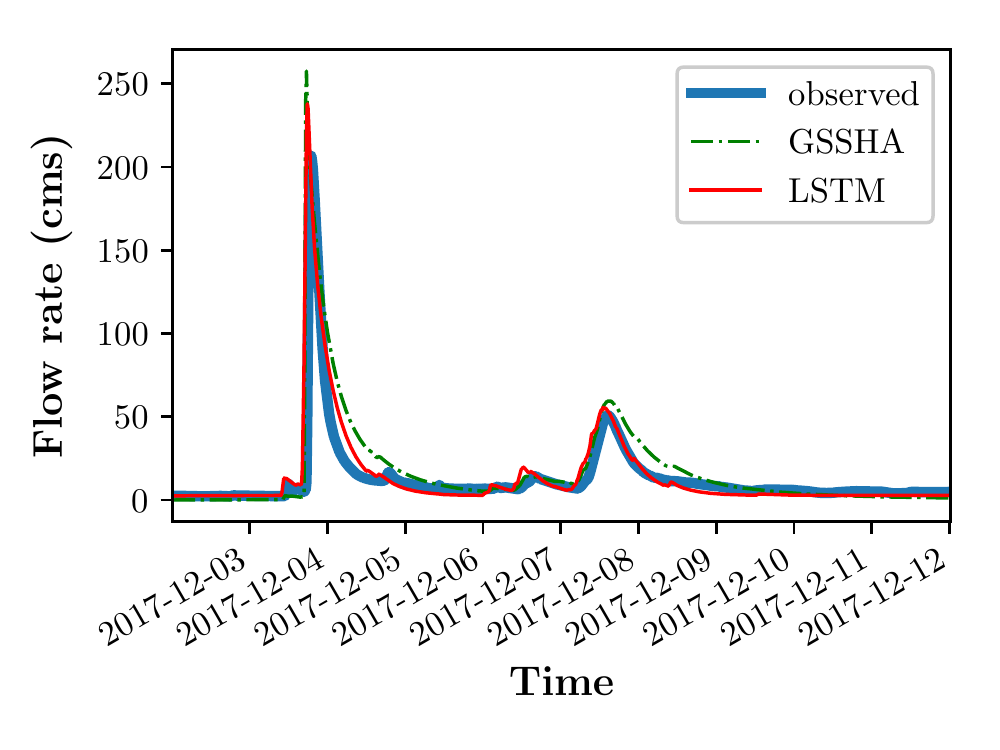}}
        \caption{High rainfall event}
        \label{fig:moderate}
     \end{subfigure}
    \begin{subfigure}{0.49\textwidth}
        \includegraphics[width=\textwidth]{{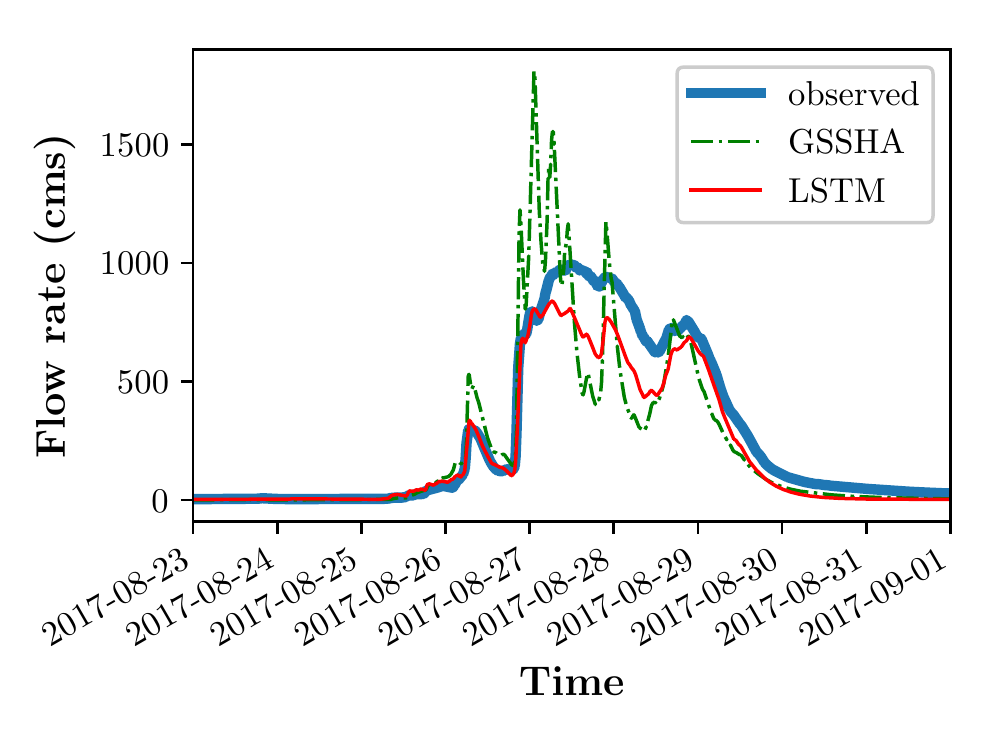}}
        \caption{Hurricane Harvey}
        \label{fig:harvey}
     \end{subfigure}
    \caption{Comparison of the ground truth flow rates and predicted flow rates computed by GSSHA and LSTM. The distinct initial gap between prediction of GSSHA model and observation shown in Figure (\ref{fig:very_low}) and (\ref{fig:low}), is due to cold starting GSSHA simulation without an initial condition. Such a gap becomes invisible as event scale increases (Figure (\ref{fig:moderate}) and (\ref{fig:harvey})).}
    \label{fig:cdq}
\end{figure}

Due to the superior performance of the 10-gage LSTM model chosen through feature selection, this model was used as the final data-driven model in this study. Thus, for the rest of this paper, the 10-gage model is referred as the LSTM model unless indicated otherwise. Figure \ref{fig:cdq} shows the comparison between the predicted flow rates by GSSHA and LSTM with the observed data. The evaluation metrics were computed and presented in Table \ref{tab:gssha-lstm-compare}. The performances of GSSHA and LSTM were similar at low or moderate rainfall events as shown in Figure \ref{fig:very_low} and \ref{fig:low}. However, when it comes to higher precipitation events, LSTM offers prediction that is closer to the observed value and is more robust compared to GSSHA.
For Hurricane Harvey, the GSSHA model over-predicted the peak flow by 82.0\%, while LSTM under-predicted the peak flow by 15.6\%. Moreover, the oscillation of GSSHA prediction depicted in Figure \ref{fig:harvey} indicates possible numerical stability issue or lower than ground truth roughness coefficient.

\begin{table}[h]
    \caption{Prediction performance comparison of GSSHA and LSTM model on selected events}
    \centering
    \begin{tabular}{l c c c c}
    \hline
    \multirow{2}{*}{Event}    & \multicolumn{2}{c}{GSSHA} & \multicolumn{2}{c}{LSTM} \\
     & RMSE & NSE & RMSE & NSE \\
    \hline
    Low event & 2.508 & 0.636 & 2.44& 0.655\\
    Moderate event & 9.00 & 0.768 & 10.02 & 0.712\\
    High event & 16.43 & 0.506 & 13.81 & 0.651\\
    Harvey & 182.74 & 0.710 & 96.64 & 0.919\\
    \hline
    \end{tabular}
    \label{tab:gssha-lstm-compare}
\end{table}

One possible solution to improve the performance of the GSSHA model on extreme events is to calibrate GSSHA with higher precipitation events or considering more calibration variables. However simulating extreme events using GSSHA is time consuming due to the stability limitations of the channel flow solver \cite{downer2000advances}. Hence, a calibration process that requires multiple forward model runs can be prohibitively expensive in this case. Another solution would be improving the quality of geometry input variables, activating the infiltration module, and increasing the model spatial resolution. All of these solutions will dramatically increase the cost of data acquisition, model set-up and computation.

\section{Conclusions}

In this study, the potential use of Long-Short-Term-Memory networks (LSTM) for RR modelling, using 15-minutes discharge and precipitation data were successfully tested for the first time. In addition, the interpretability of the LSTM model in terms of its attention distribution on the input space was explored.

The designed numerical tests showed that the LSTM model can accurately predict stream flow given precipitation as the sole input when the scales of test data and training data are identical (interpolation). Furthermore, an LSTM model with the proper regularization and choice of rainfall gages can even extrapolate well as shown in the prediction of Hurricane Harvey.
The performance of the LSTM model was remarkable considering the training data was highly skewed towards base flow rate when little precipitation was presented.
As complicated as the LSTM model is, the gage importance can be evaluated using the weights defined in this study. This is mainly because the gage readings are measures of the same physics quantity and have similar scale and distribution. 

When compared to the process-driven model, GSSHA, the data-driven model is clearly more efficient and robust in terms of prediction and calibration. A 20-day event with extreme precipitation can be predicted at a click of the mouse using LSTM, while the same prediction using GSSHA can take hours to a day depending on the topography of the watershed and resolution of the discretization. The results of this study also showed that the calibration of the LSTM network using the 10-years data set is still faster than calibrating GSSHA using a 20-days event including Hurricane Harvey. Besides, given that both the data-driven and the process-driven models were not extensively tuned, the LSTM model is more robust and accurate while predicting high precipitation events. Calibration of the process-driven models would be restricted by its prediction efficiency while data-driven models do not have such restrictions.

\section{Acknowledgement}
This research was funded by the Severe Storm Prediction, Education and Evacuation from Disasters Center [grant number R09252] and the National Oceanic and Atmospheric Administration [grant number NA18NOS0120158]. Their support is gratefully acknowledged. This is a pre-print of an article published in Neural Computing and Applications. The final authenticated version is available online at: https://doi.org/10.1007/s00521-020-05010-6.

\bibliographystyle{unsrt}  
\bibliography{main}  

\newpage
\appendix
\renewcommand{\thefigure}{A.\arabic{figure}}

\setcounter{figure}{0}

\setcounter{table}{0}
\renewcommand{\thetable}{A.\arabic{table}}
\section{Supplementary Information} \label{SI}

\begin{figure}[H]
    \centering
    \includegraphics[width=0.9\linewidth]{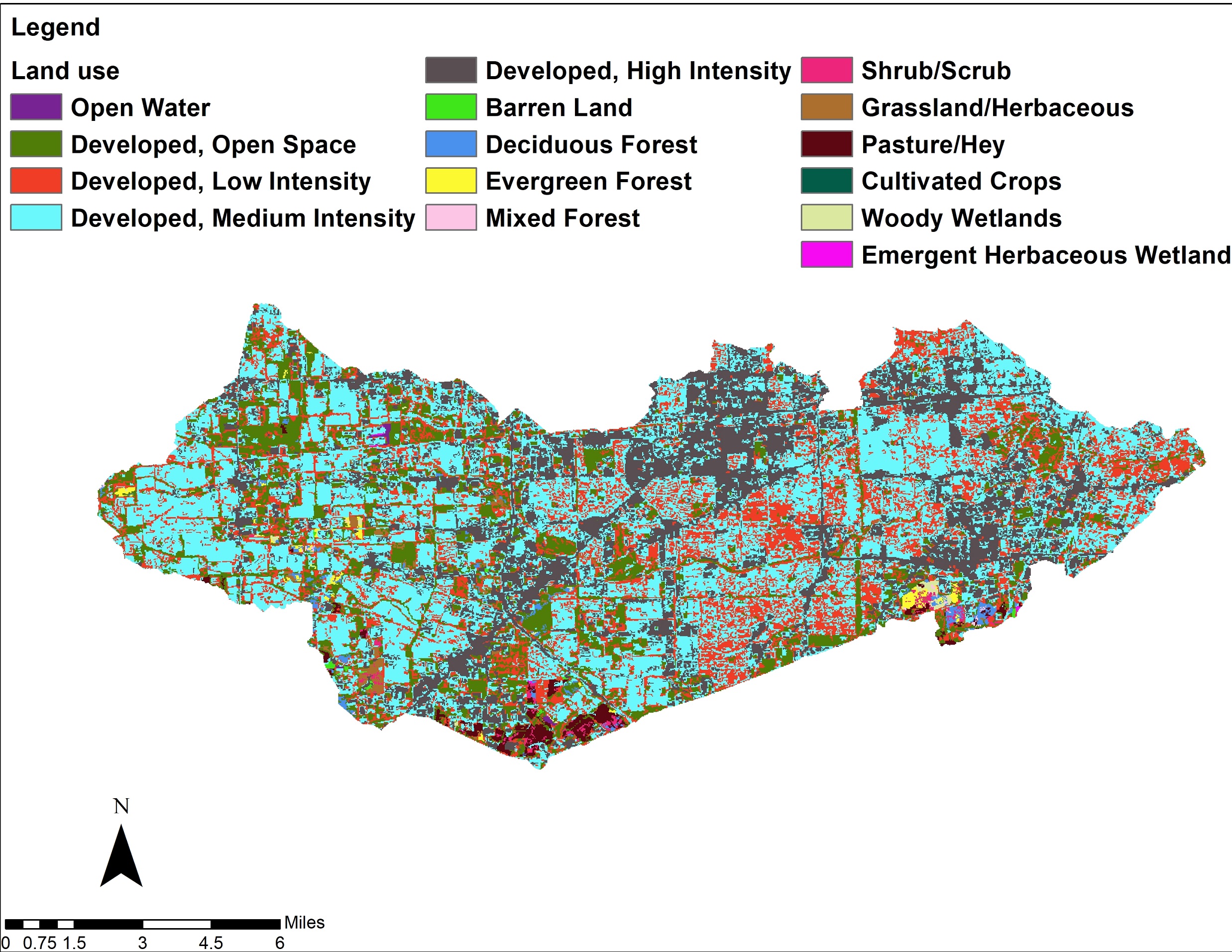}
    \caption{Land use of study area}
    \label{fig:land_use}
\end{figure}

\begin{table}[H]
    \centering
    \begin{tabular}{c c c}
    \hline
    NLCD Value & Normal Manning's n value & Allowable range of n values \\
    \hline
    11 & 0.040 & 0.025-0.05 \\
    21 & 0.040 & 0.03-0.05 \\
    22 & 0.100 & 0.08-0.12 \\
    23 & 0.080 & 0.06-0.14 \\
    24 & 0.150 & 0.12‐0.20 \\
    31 & 0.025 & 0.023-0.030 \\
    41 & 0.160 & 0.10-0.16 \\
    42 & 0.160 & 0.10-0.16 \\
    43 & 0.160 & 0.10-0.16 \\
    52 & 0.100 & 0.07-0.16 \\
    71 & 0.035 & 0.025-0.050 \\
    81 & 0.030 & 0.025-0.050 \\
    82 & 0.035 & 0.025-0.050 \\
    90 & 0.120 & 0.045-0.15 \\
    95 & 0.070 & 0.05-0.085 \\
    \hline
    \end{tabular}
    \caption{Manning’s n values for various land covers to use for GSSHA simulation.}
    \label{tab:land_use}
\end{table}

\begin{figure}[H]
    \centering
    \begin{subfigure}{0.49\textwidth}
        \includegraphics[width=\textwidth]{{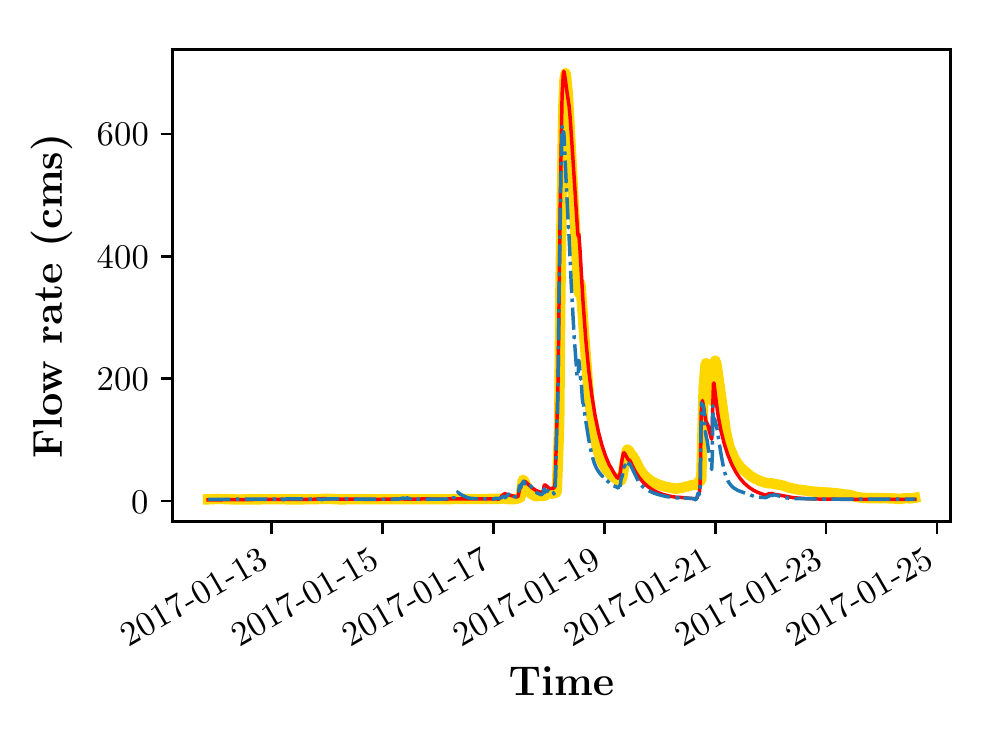}}
     \end{subfigure}
     \begin{subfigure}{0.49\textwidth}
        \includegraphics[width=\textwidth]{{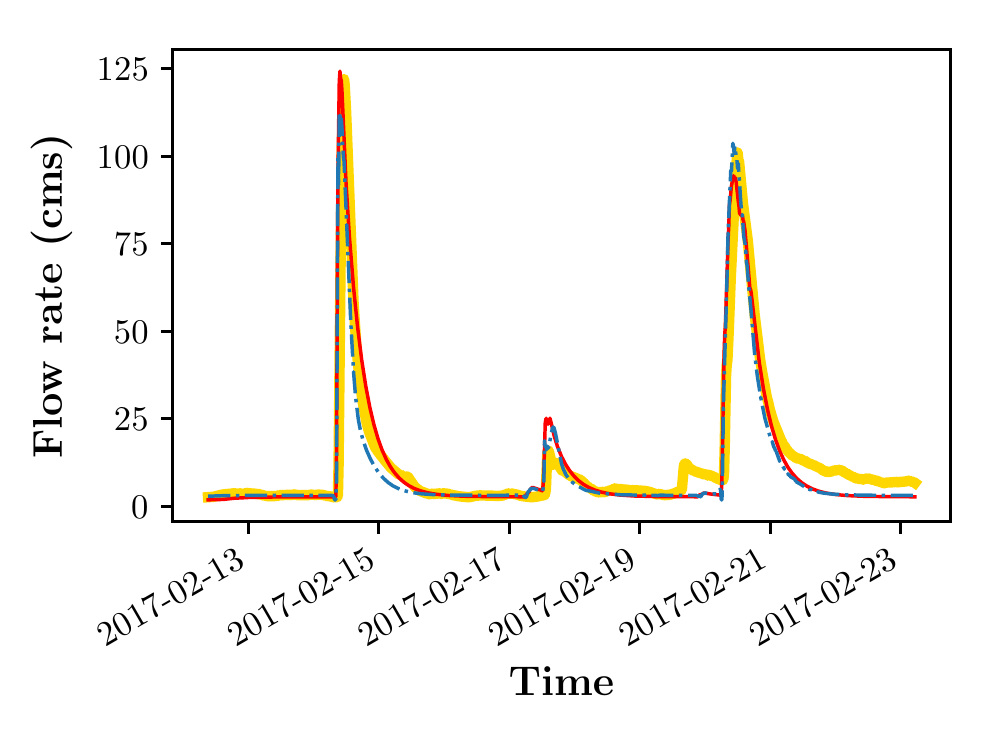}}
     \end{subfigure}
     \begin{subfigure}{0.49\textwidth}
        \includegraphics[width=\textwidth]{{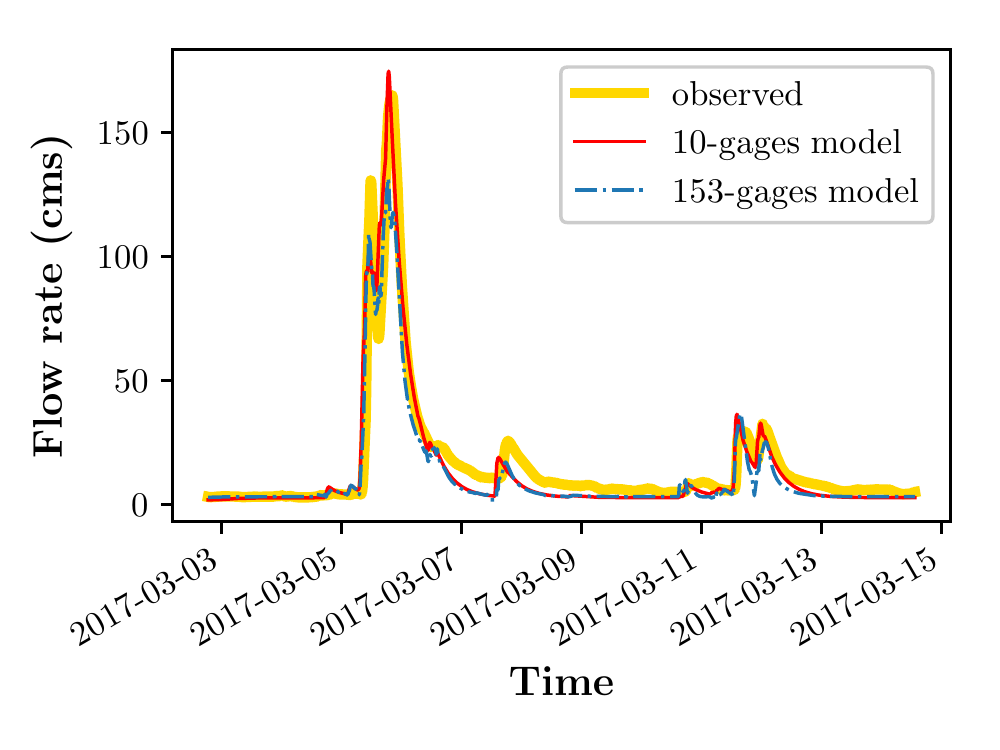}}
     \end{subfigure}
    \begin{subfigure}{0.49\textwidth}
        \includegraphics[width=\textwidth]{{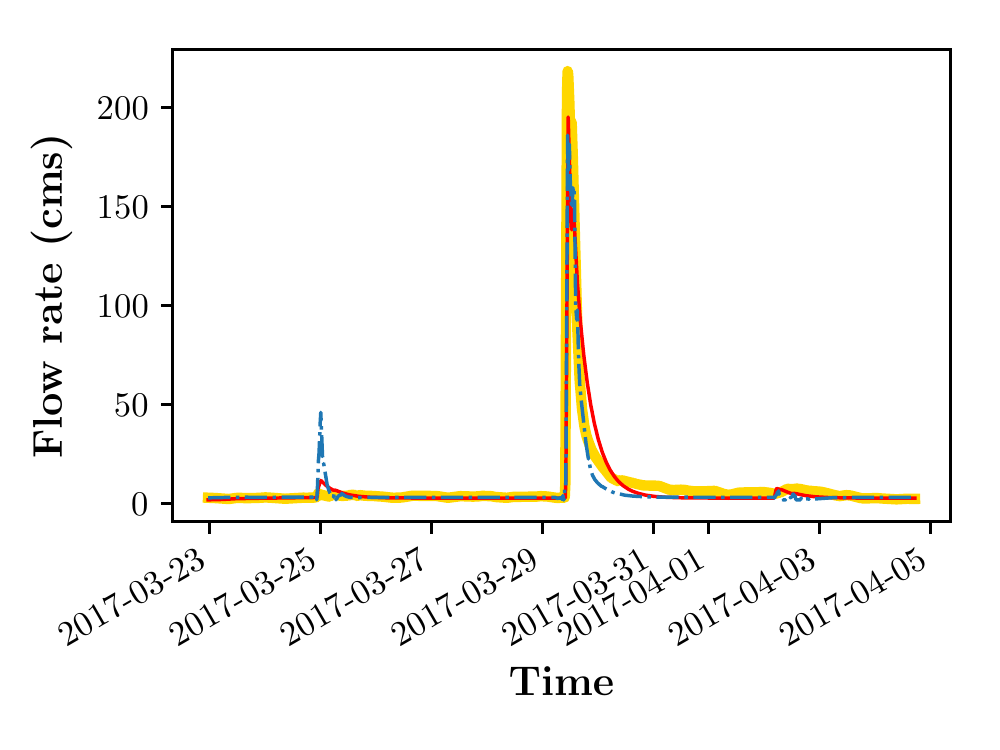}}
     \end{subfigure}
     \begin{subfigure}{0.49\textwidth}
        \includegraphics[width=\textwidth]{{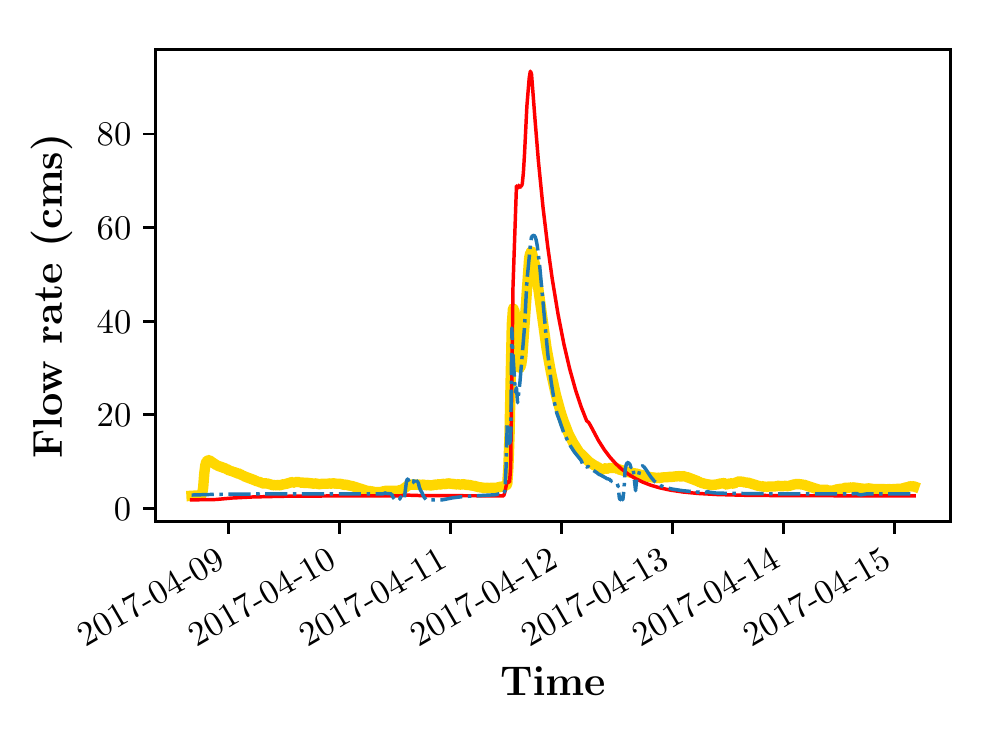}}
     \end{subfigure}
    \begin{subfigure}{0.49\textwidth}
        \includegraphics[width=\textwidth]{{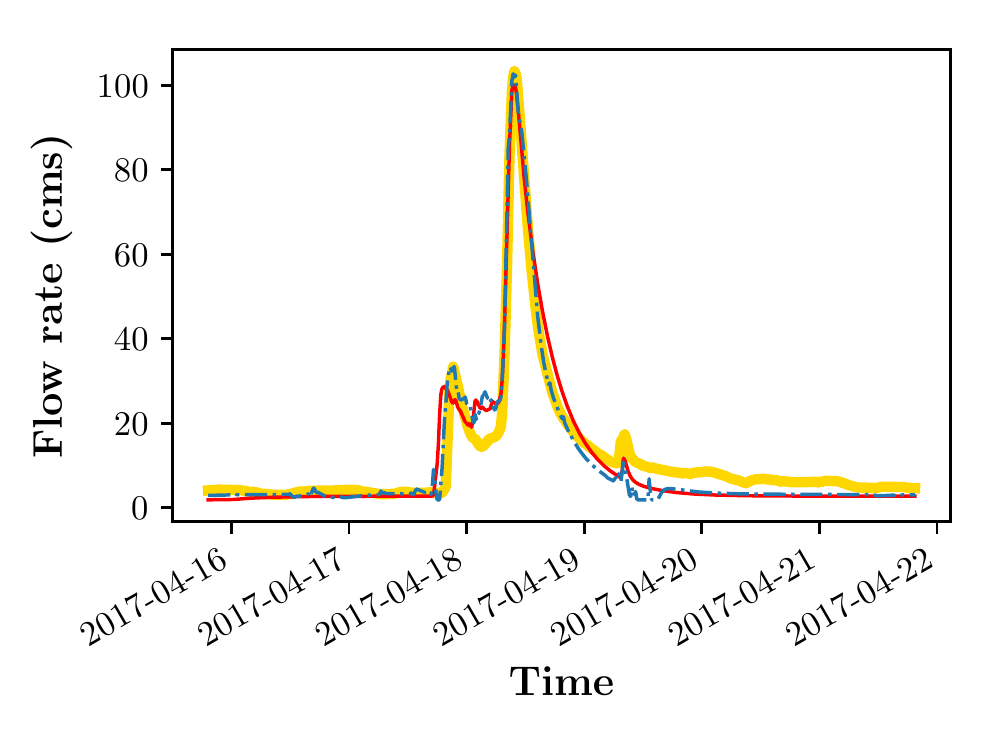}}
     \end{subfigure}
    \caption{Comparison of observation, 10-gages LSTM, and 153-gages LSTM}
    \label{fig:lstm-gage-test-all}
\end{figure}
\begin{figure}[H]\ContinuedFloat
    \centering
    \centering
    \begin{subfigure}{0.49\textwidth}
        \includegraphics[width=\textwidth]{{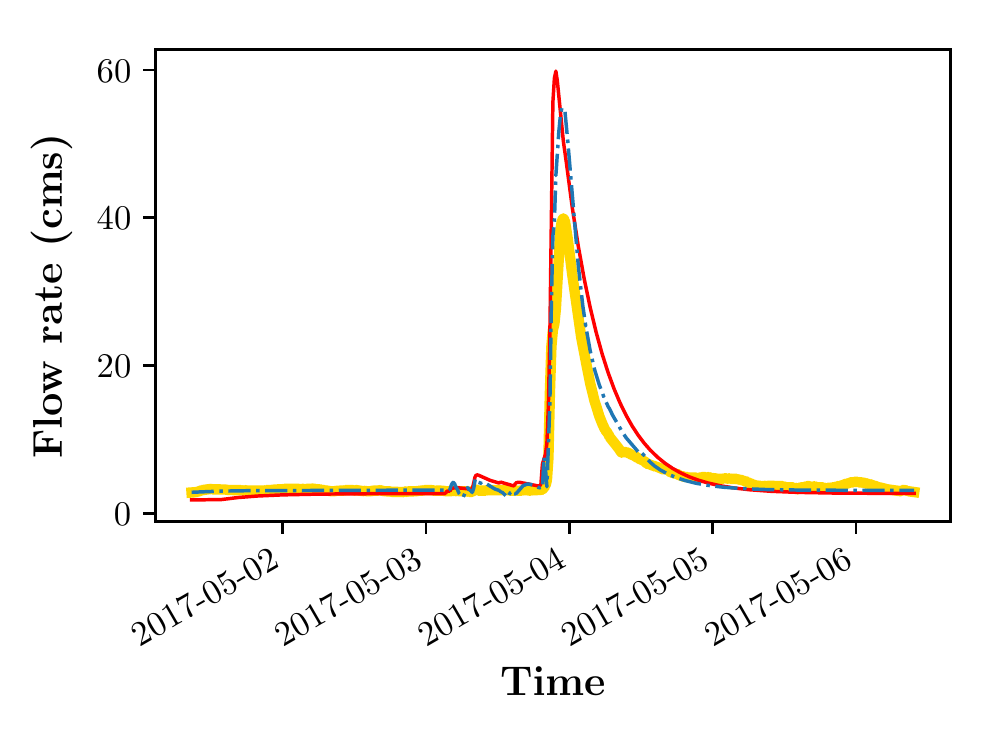}}
     \end{subfigure}
     \begin{subfigure}{0.49\textwidth}
        \includegraphics[width=\textwidth]{{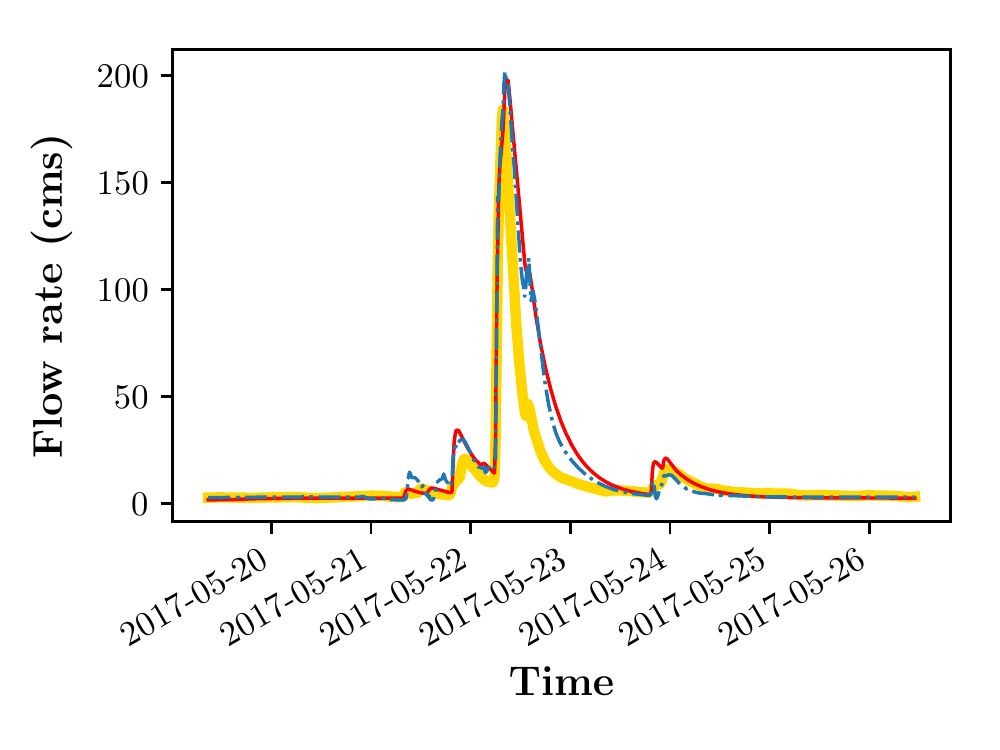}}
     \end{subfigure}
     \begin{subfigure}{0.49\textwidth}
        \includegraphics[width=\textwidth]{{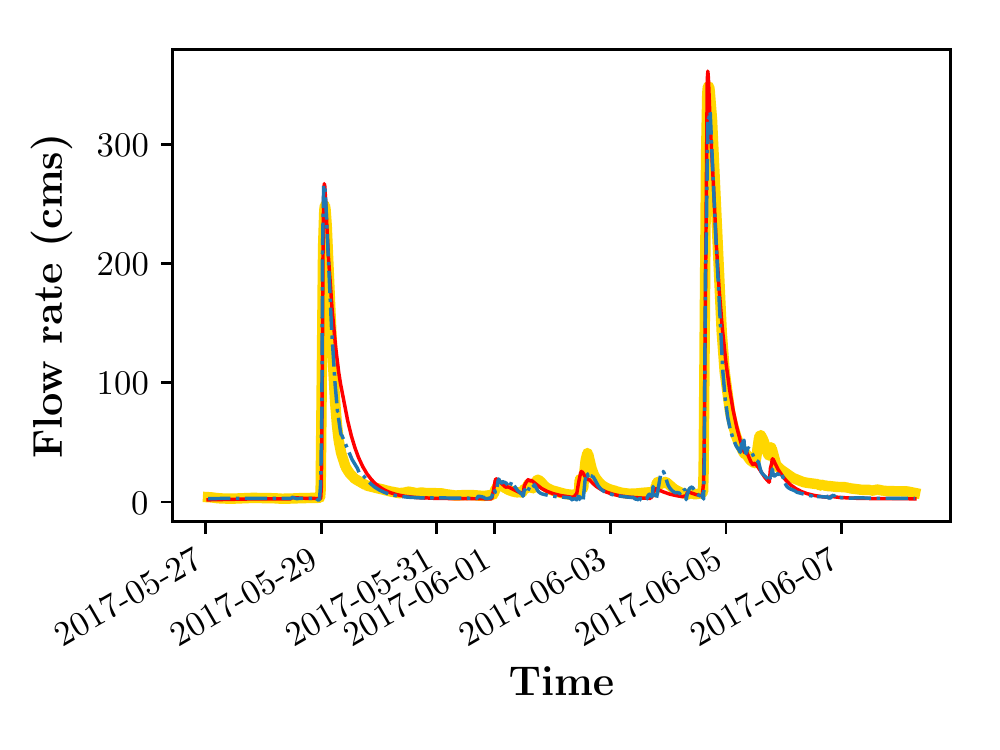}}
     \end{subfigure}
    \begin{subfigure}{0.49\textwidth}
        \includegraphics[width=\textwidth]{{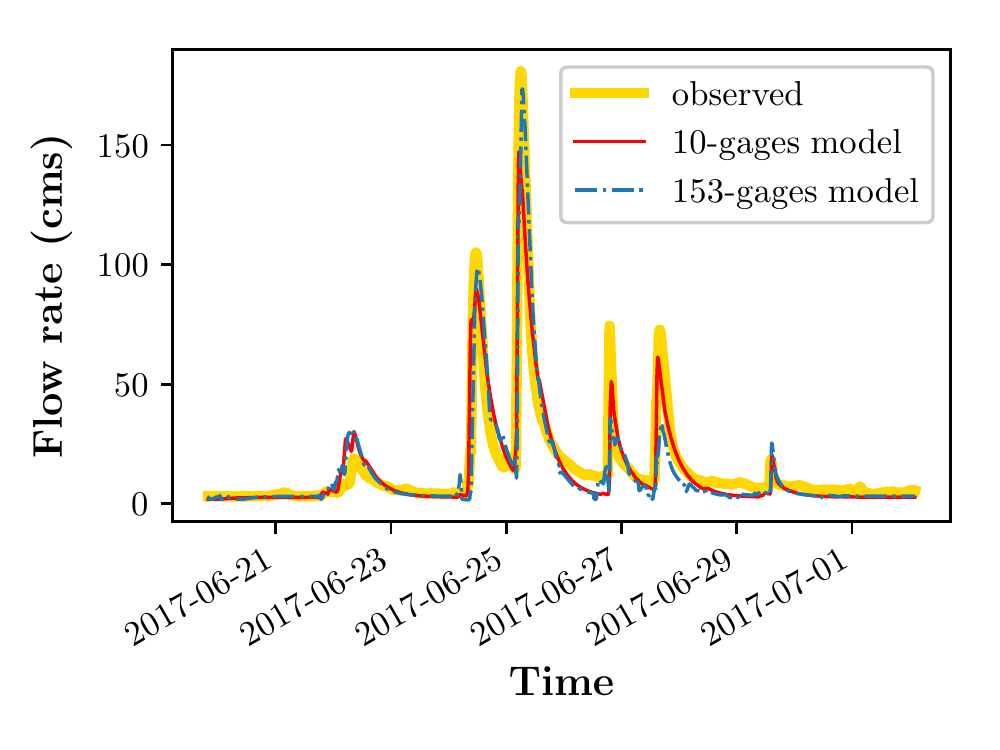}}
     \end{subfigure}
     \begin{subfigure}{0.49\textwidth}
        \includegraphics[width=\textwidth]{{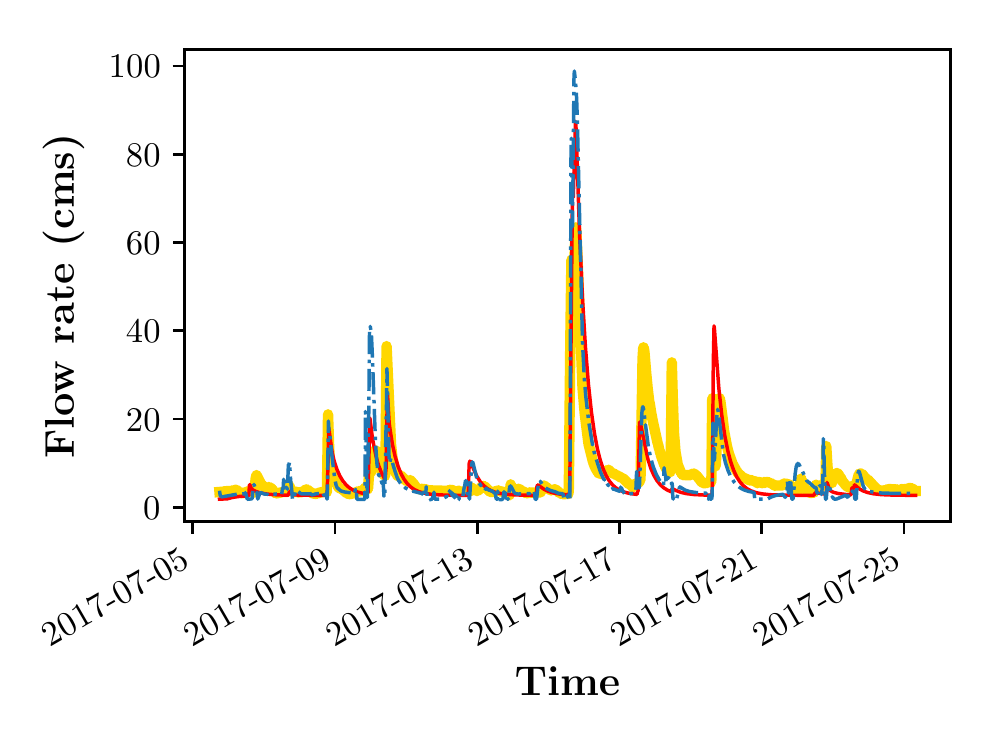}}
     \end{subfigure}
    \begin{subfigure}{0.49\textwidth}
        \includegraphics[width=\textwidth]{{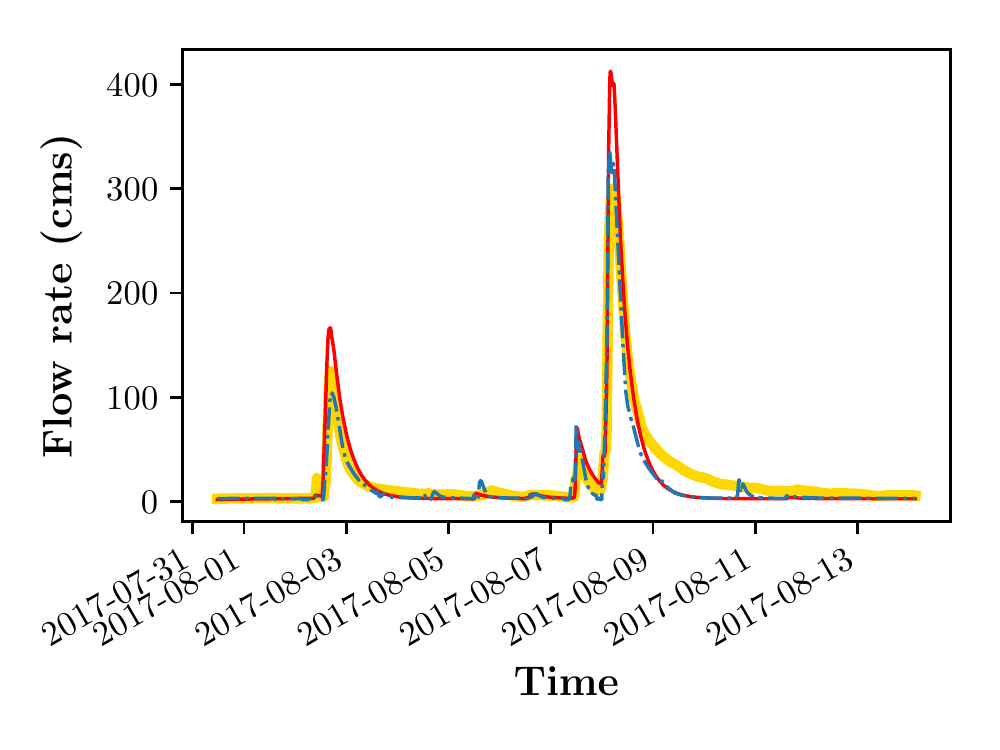}}
     \end{subfigure}
    \caption{(Cont.) Comparison of observation, 10-gages LSTM, and 153-gages LSTM}
\end{figure}
\begin{figure}[H]\ContinuedFloat
    \centering
    \begin{subfigure}{0.49\textwidth}
        \includegraphics[width=\textwidth]{{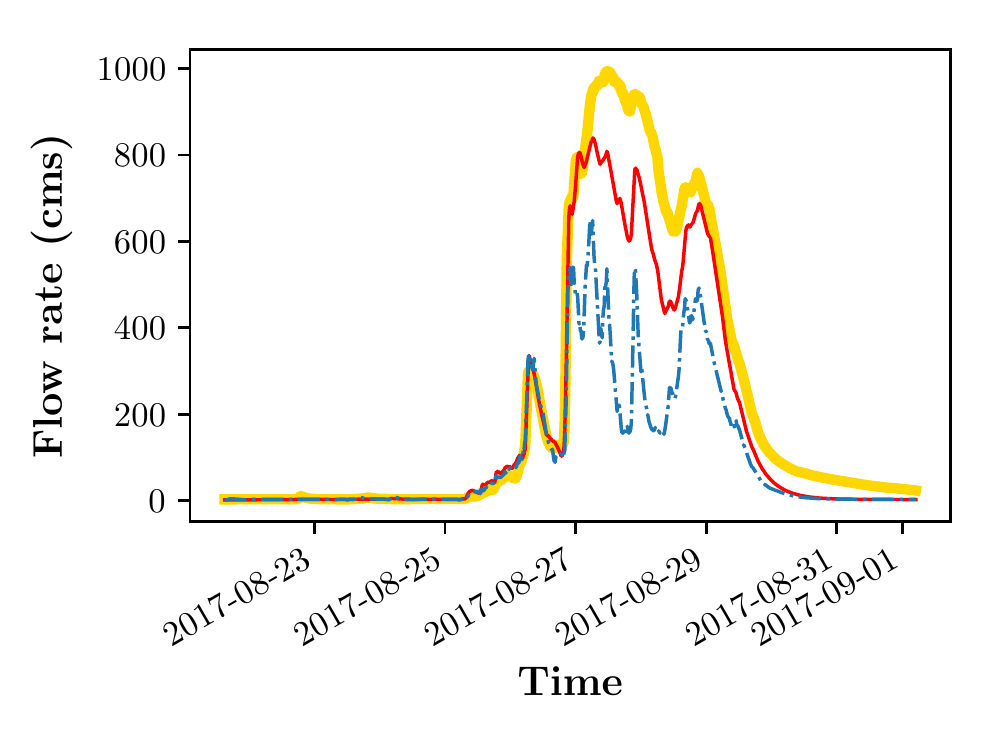}}
     \end{subfigure}
     \begin{subfigure}{0.49\textwidth}
        \includegraphics[width=\textwidth]{{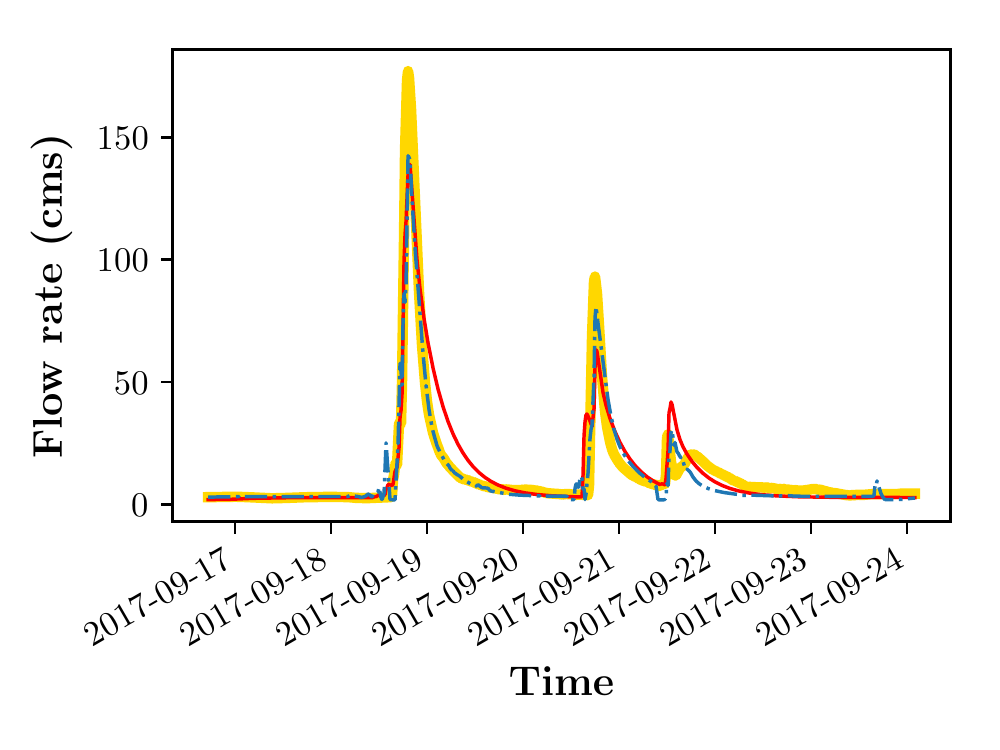}}
     \end{subfigure}
     \begin{subfigure}{0.49\textwidth}
        \includegraphics[width=\textwidth]{{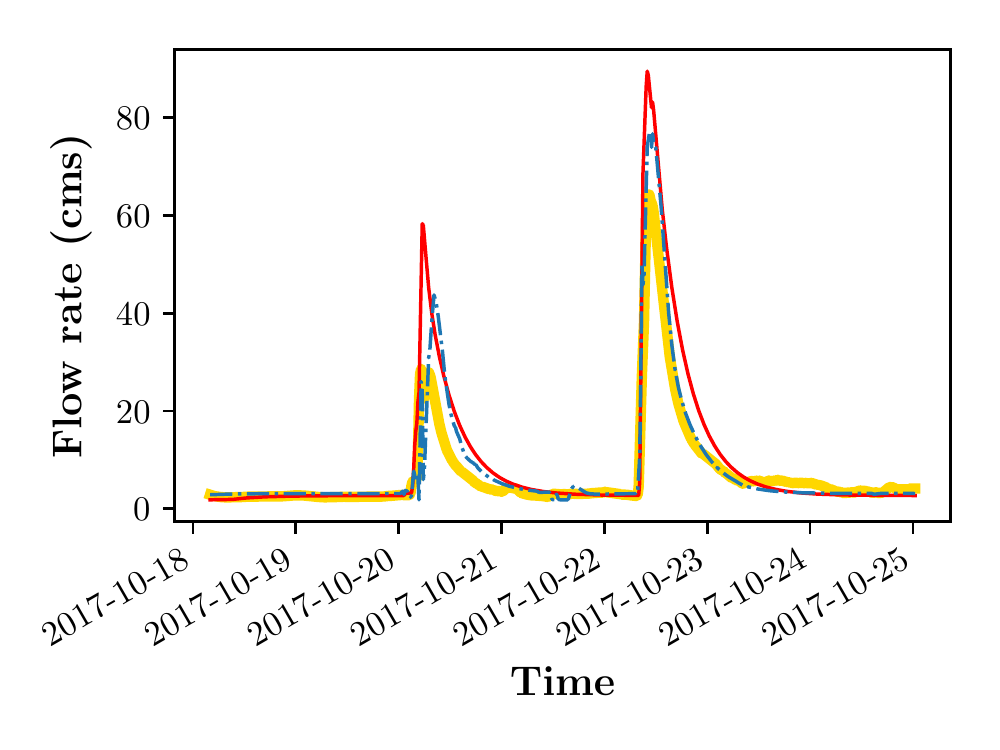}}
     \end{subfigure}
    \begin{subfigure}{0.49\textwidth}
        \includegraphics[width=\textwidth]{{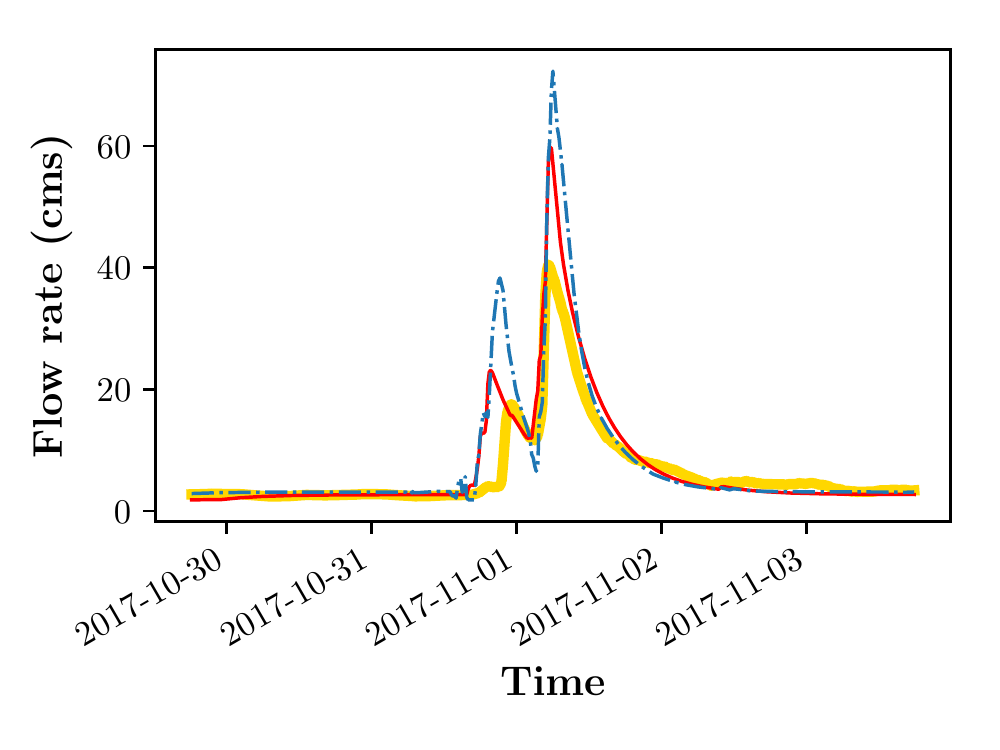}}
     \end{subfigure}
     \begin{subfigure}{0.49\textwidth}
        \includegraphics[width=\textwidth]{{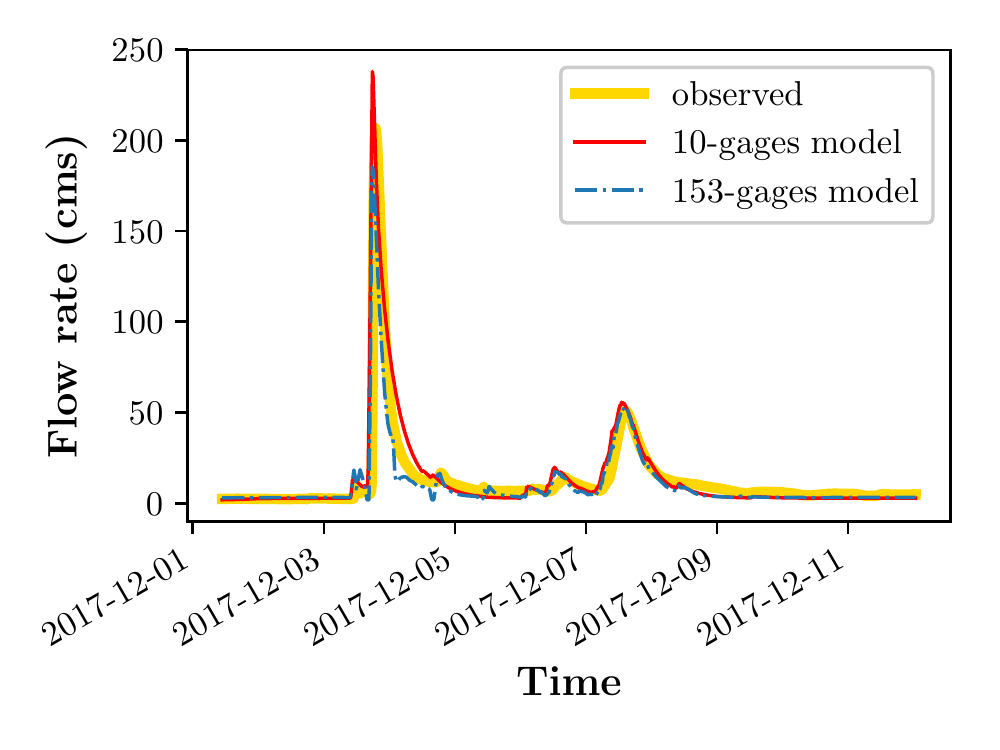}}
     \end{subfigure}
    \begin{subfigure}{0.49\textwidth}
        \includegraphics[width=\textwidth]{{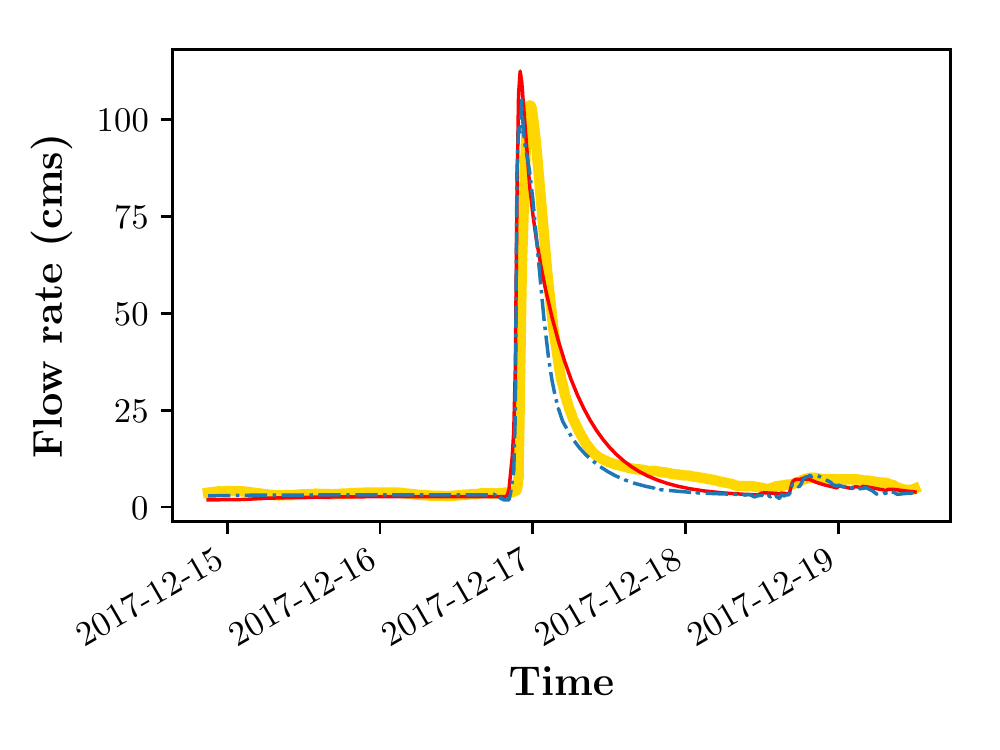}}
     \end{subfigure}
    \caption{(Cont.) Comparison of observation, 10-gages LSTM, and 153-gages LSTM. A close look at the lower discharge region would reveal the unphyscial oscillation of 153-gages prediction, indicating overfitting.}
\end{figure}

\end{document}